\documentclass[twocolumn,showpacs,superscriptaddress]{revtex4}

\usepackage{amsmath}
\usepackage{bm}
\usepackage{array}
\usepackage{graphicx}
\usepackage{dcolumn}   % Align table columns on decimal point
\usepackage{amssymb}
\usepackage[english]{babel}

\begin{document}

%\preprint{PRL}
\preprint{PRA}

\title{Perturbative behaviour of a vortex in a trapped Bose-Einstein condensate}
\date{\today}

\author{Lyndon Koens}
\affiliation{School of Physics, The University of Melbourne, Parkville,
3010, Australia}
\author{Andrew M. Martin}
\affiliation{School of Physics, The University of Melbourne, Parkville,
3010, Australia}

\begin{abstract}
We derive a set of equations that describe the shape and behaviour of a single perturbed vortex line in a Bose-Einstein condensate. Through the use of a matched asymptotic expansion and a unique coordinate transform a relation for a vortex's velocity, anywhere along the line, is found in terms of the trapping, rotation, and distortion of the line at that location. This relation is then used to find a set of differential equations that give the line's specific shape and motion. This work corrects a previous similar derivation by Anatoly A. Svidzinsky and Alexander L. Fetter [Phys. Rev. A \textbf{62}, 063617 (2000)], and enables a comparison with recent numerical results.
\end{abstract}
\maketitle
\section{Introduction}

Experimental realisations of vortices in Bose-Einstein condensates (BECs) \cite{PhysRevLett.83.2498, PhysRevLett.85.2857, PhysRevLett.84.806, PhysRevLett.85.2223, PhysRevLett.92.050403} has spurred lots of research into the behaviour and effects vortices have on a BEC \cite{5502338220101201,RevModPhys.81.647}. Similar to the vortices Feynman predicted in superfluid He II \cite{Prog.LowTemp.Phys.1.17.}, vortices in BECs have quantised circulation, ensuring that the condensates phase is a single valued function (see Ref.~\cite{PethickSmith} for details). However unlike in superfluid He II, the vortices in a BEC are sufficiently described in the mean field regime by the Gross-Pitaevskii equation \cite{Sov.Phys.JETP.13.451, NuovoCimento.20.454} with rotation [Eq.~\eqref{gp}].

\begin{gather} 
  \left[ -\frac{\hbar^{2}}{2 M} \nabla^{2} + V_{tr}(\mathbf{r}) + g |\Psi|^{2}  - \mu(\Omega)\right]\Psi \nonumber  \\
  + i \hbar \mathbf{\Omega}\cdot \left(\mathbf{r}\times\nabla\right)\Psi  = i \hbar \frac{\partial \Psi}{\partial t}, \label{gp}
 \end{gather}
where $\Psi$ is the condensate wavefunction, $V_{tr}(\mathbf{r})$ is the external trapping potential, $\mathbf{\Omega}$ is the condensate rotation vector, $g = 4 \pi a \hbar^{2} /M$ is the interparticle interaction strength, $\mu(\Omega)$ is the chemical potential, $M$ is the particle mass, and $a$ is the s-wave scattering length.

In a stationary condensate it is energetically unfavourable to contain vortices, however when $\Omega > \Omega_{c} =5\hbar \ln\left(R_{\perp}/r_{c}\right)/2 M R_{\perp}^{2}$ ($R_{\perp}$ being the radius of the BEC perpendicular to the rotation vector and $r_{c}$ being the vortex core radius) a centred vortex becomes energetically favourable \cite{PhysRevLett.84.5919}. Upon investigating the dissipation of such a vortex, the vortex's dissipation time was shown to depend on the condensate's temperature \cite{PhysRevA.60.R1779}.

In a trapped BEC a quantised vortex is rarely stationary, usually moving or contorting. An off centred straight line vortex is known to precess at $\dot{\phi} = 3 \Omega_{c} / 5(1- \rho_{0}^{2}/R_{\perp}^{2})$  ($\rho_{0}$ being the cylindrical radial coordinate of the vortex line) \cite{PhysRevLett.84.5919, PhysRevA.63.043608}. This straight structure is representative of a vortex in a pancake shape condensate, $R_{\perp}/R_{z} > 1$, where $R_{z}$ the condensate radius parallel to the rotation vector. For a cigar shaped trap, $R_{\perp}/R_{z} < 1$, however the vortex line bends. This result was shown numerically in Refs.~\cite{PhysRevA.64.043611,PhysRevA.64.053611}, seen experimentally by Rosenbusch \textit{et al.} \cite{PhysRevLett.89.200403}, and reasoned physically by M. Modugno \textit{et al.} \cite{Eur.Phys.J.D.22}. A quantised vortex also supports wave perturbations, coupling with and modifying the normal modes of the condensate \cite{PhysRevLett.81.1754, PhysRevLett.86.2922} or supporting helical wave structures, like Kelvin waves \cite{Philos.mag.10.155}, along its length \cite{Sov.Phys.JETP.13.451, PhysRevLett.90.100403}.

With such a wide range of behaviour, a single set of equations that completely describes the vortex has been desired. Early attempts to derive this equation relied on the method of matched asymptotic expansion \cite{PhysicaD78.1994.1, PhysicaD.47.1991.353}. This procedure was extended by A. Svidzinsky and A. Fetter \cite{Phys.Rev.A.62.063617} to produce a set of equations that should generally described all small perturbations from a straight line vortex. This derivation used a coordinate transform to account for vortex line bending, however in its execution, these coordinates were not formally defined, and the outer and inner solutions matched through a pseudo `vector potential'. This gave rise to a modified set of equations to those that describe the vortex's motion and structure.

This paper re-derives this procedure to determine the behaviour and structure of a slightly perturbed straight line vortex, using coordinate transformations and the method of match asymptotic expansion. Section \ref{sec:idea} justifies the relevance and procedure of such an expansion, and Section \ref{sec:Vclc} defines the unique coordinate system used. An inner and outer solution are then found (Sections \ref{sec:inner} and \ref{sec:outer} respectively) and matched (Section \ref{sec:match}) to give a relation between the perturbed vortex's motion and shape. This relation contains an unknown constant that is determined by comparing to known physical scenarios (Section \ref{sec:DE}). This then allows simple results from the relation to be calculated and compared to numerical simulations (Section \ref{sec:waves}). Finally in Section \ref{sec:Psw}, a general comparison between this work and Ref. \cite{Phys.Rev.A.62.063617} is performed. Illustrating the differences in methodology and general results.

\section{The Idea} \label{sec:idea} 
BECs with vortices within them have two natural length scales: the condensate length scale $R_{\perp}$ and the vortex core radius $r_{c}$, with $r_{c} < R_{\perp}$. These two scales suggest that a matched asymptotic expansion \cite{Verhulst} can be performed to determine the behaviour near the vortex core, length scale $r_{c}$, far from the vortex core, length scale $R_{\perp}$, and matched to give the systems full behaviour. Such an asymptotic expansion has been previously preformed for vortices in simplified scenarios in Refs.~\cite{PhysicaD78.1994.1} and \cite{PhysicaD.47.1991.353}. 

In the local coordinates of the vortex line the inner solution depends on properties of the vortex line, which when matched to the outer solution creates a relation between the shape and velocity of the line. This relationship then gives a set of differential equations that describes the vortex line's structure and motion through space.

\section{The local coordinates of the vortex line} \label{sec:Vclc}

In a BEC the localized vorticity from a vortex, creates a single vortex line. Each point on this line has a position and a velocity; therefore, the vortex line, at a given time, is described by two parametrized functions: a curve describing the lines shape and position, $\mathbf{c}(T)$, and a velocity vector, $\mathbf{V}(T)$. This parametrization allows for the transformation into the local coordinate system; the curve, $\mathbf{c}(T)$, determining the axes and $\mathbf{V}(T)$ representing the line's motion at a given point.

\begin{figure}[ht]
\centering
\includegraphics[width=0.4\textwidth]{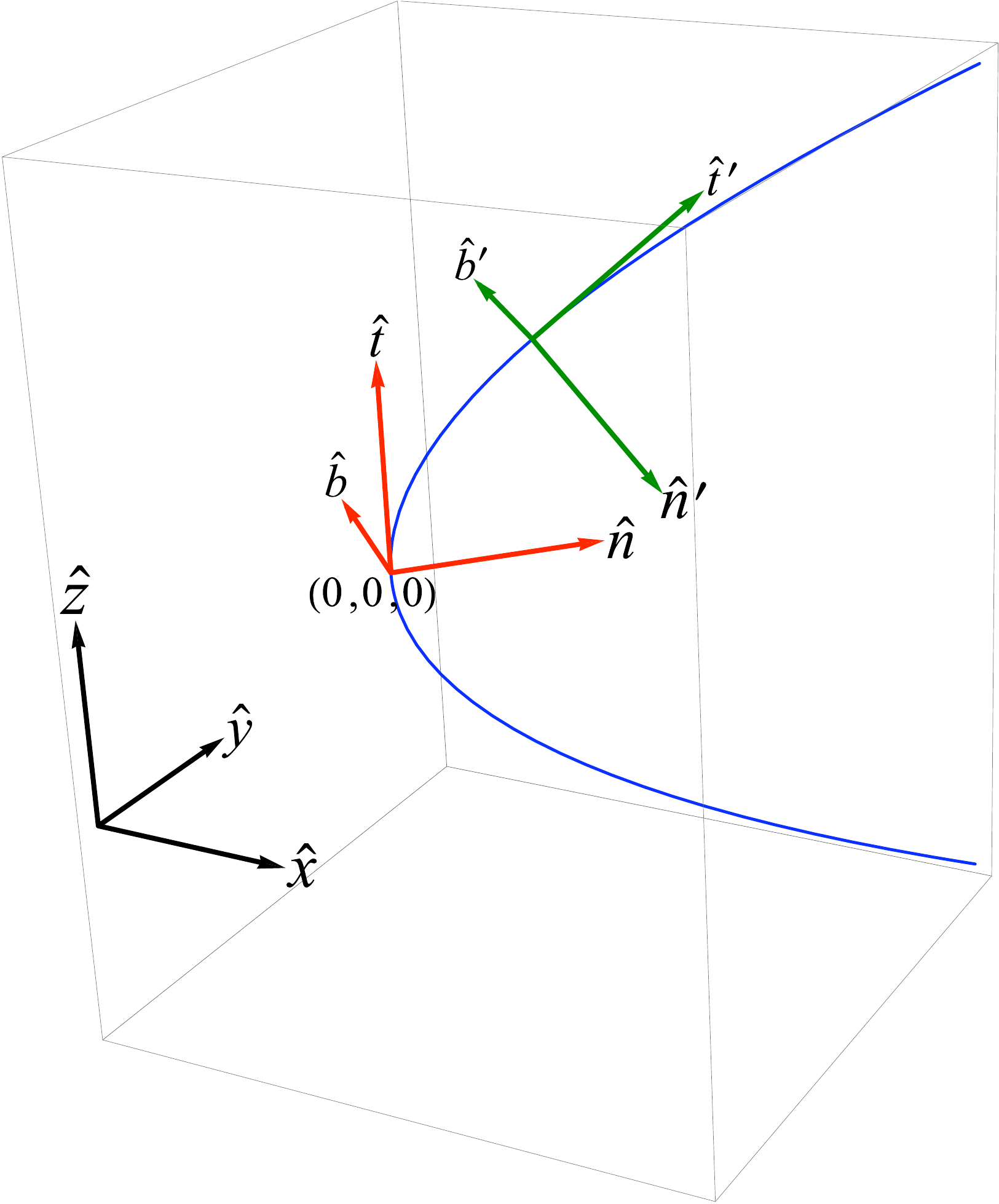}
\caption{ Depiction of the local coordinate vectors $\mathbf{\hat{t}}$, $\mathbf{\hat{n}}$ and $\mathbf{\hat{b}}$ of a curve $\mathbf{c}(T) = \{T^{2}/2,T^{2}/2,T\}$ (blue), with the coordinate vectors plotted for $T=0$ (red) and $T=1$ (green). }
\label{fig:coords1}
\end{figure}

For any curve [$\mathbf{c}(T)$], a tangent vector [Eq.~\eqref{tangentv}], a normal vector [Eq.~\eqref{normalv}], and a binormal vector [Eq.~\eqref{binormalv}] are defined, see Fig.~\ref{fig:coords1}. These three orthogonal vectors form the basis to the coordinate system of the curve, known as the Frenet-Serret coordinates \cite{KühnelWolfgang},
\begin{eqnarray}
\mathbf{\hat{t}} &=& \frac{\mathbf{c}'(T)}{\left|\mathbf{c}'(T)\right|}, \label{tangentv} \\
\mathbf{\hat{n}} &=& \frac{\mathbf{c}'(T) \times(\mathbf{c}''(T)\times \mathbf{c}'(T))}{\left|\mathbf{c}'(T)\right| \left|\mathbf{c}''(T)\times \mathbf{c}'(T)\right|}, \label{normalv} \\
\mathbf{\hat{b}} &=& \mathbf{\hat{t}} \times \mathbf{\hat{n}}. \label{binormalv}
\end{eqnarray}

At a given $T$, such a basis is equivalent to the basis of a Cartesian coordinate system. Allowing the respective $\mathbf{\hat{t}}$, $\mathbf{\hat{n}}$, and $\mathbf{\hat{b}}$ vectors to be locally treated as $\mathbf{\hat{z}}$, $\mathbf{\hat{x}}$, and $\mathbf{\hat{y}}$.

This local Cartesian behaviour extends to the derivatives in $\mathbf{\hat{n}}$ and $\mathbf{\hat{b}}$ directions. Translations in these directions do not change the basis vectors, and therefore the local derivatives in $\mathbf{\hat{n}}$ and $\mathbf{\hat{b}}$ are identical to their Cartesian counterparts ($\partial_{n} \equiv  \partial_{x},\partial_{n}^{2} \equiv  \partial_{x}^{2}$, etc.). 

The curve progresses in $\mathbf{\hat{t}}$ causing the local derivatives in $\mathbf{\hat{t}}$ to not match its Cartesian equivalent. Using the chain rule, this deviation can be quantified, showing the first order derivatives in $\mathbf{\hat{t}}$ to be
\begin{equation}
\left. \frac{\partial}{\partial T} \right|_{T=T_{0}} = \lim_{T\rightarrow T_{0}} \mathbf{c}'(T) \cdot \nabla,
\end{equation}
from which higher order derivatives can be constructed.
\begin{multline}
\left. \frac{\partial}{\partial T} \left(\frac{\partial}{\partial T}\right) \right|_{T=T_{0}} = \\
 \lim_{T\rightarrow T_{0}} \left[ \mathbf{c}''(T)\cdot \nabla + \mathbf{c}'(T) \otimes\mathbf{c}'(T) : \nabla \otimes\nabla \right],
\end{multline}
where $\otimes$ represents the outer product and $:$ represents a double contraction.

In this case the structure of the vortex line is unknown. Therefore in order to evaluate these derivatives a pseudo-parametrization, that describes the local behaviour and can take any shape, needs to be used.

Conveniently, any unique curve is defined through two terms: curvature [Eq.~\eqref{curvature}] and torsion [Eq.~\eqref{torsion}]. These terms describe how a curve bends and distorts through space, irrespective of its location and choice of parametrization. Curvature indicates how the curve bends in the $\mathbf{\hat{t}}$, $\mathbf{\hat{n}}$ plane (with $1/\kappa$ being the radius of a circle at a given point) and torsion indicates how the curve twists out of the $\mathbf{\hat{t}}$, $\mathbf{\hat{n}}$ plane (how $\mathbf{\hat{n}}$ and $\mathbf{\hat{b}}$ rotates as $T$ progresses): 

\begin{eqnarray}
 \kappa  &=&  \frac{\left| \mathbf{c}'(T) \times \mathbf{c}''(T) \right|}{\left|\mathbf{c}'(T)\right|^{3}}, \label{curvature} \\
 \tau &=& \frac{( \mathbf{c}'(T) \times \mathbf{c}''(T))\cdot \mathbf{c}'''(T)}{\left| \mathbf{c}'(T) \times \mathbf{c}''(T) \right|^{2}}. \label{torsion}
\end{eqnarray}

Generally, a curve's torsion and curvature vary along its length, allowing for all possible 3D curve structures to form. However, locally around any point the torsion and curvature is effectively constant, prompting the use of a modified helix %(helices have constant curvature and torsion) 
for the pseudo-parametrization of the vortex line:
\begin{equation}
\mathbf{c}(T) = \left\{ a\textbf{ } \cos(T), a\textbf{ } \sin(T), b\textbf{ } T \right\} .
\end{equation}

Assuming the excitations on a vortex line are small compared with the vortex's overall structure the curves tangent vector will almost align with $\mathbf{\hat{z}}$. This condition can be enforced by adding appropriate linear terms to the parametrization, to make $\mathbf{\hat{t}} \vert_{T=0} \approx \mathbf{\hat{z}}$. This condition also implies that the length of the curve is predominantly the length travelled in $z$ ($b$ is large); this simplifies the re-normalization of the parametrization, from $T$ into the arc length of the curve $s \approx z =  b T$. This is an easier parametrization to work with as curves parametrized by arc length have $| \mathbf{c}'(s) |$  = 1.

\begin{figure}[ht]
\centering
\includegraphics[width=0.4\textwidth]{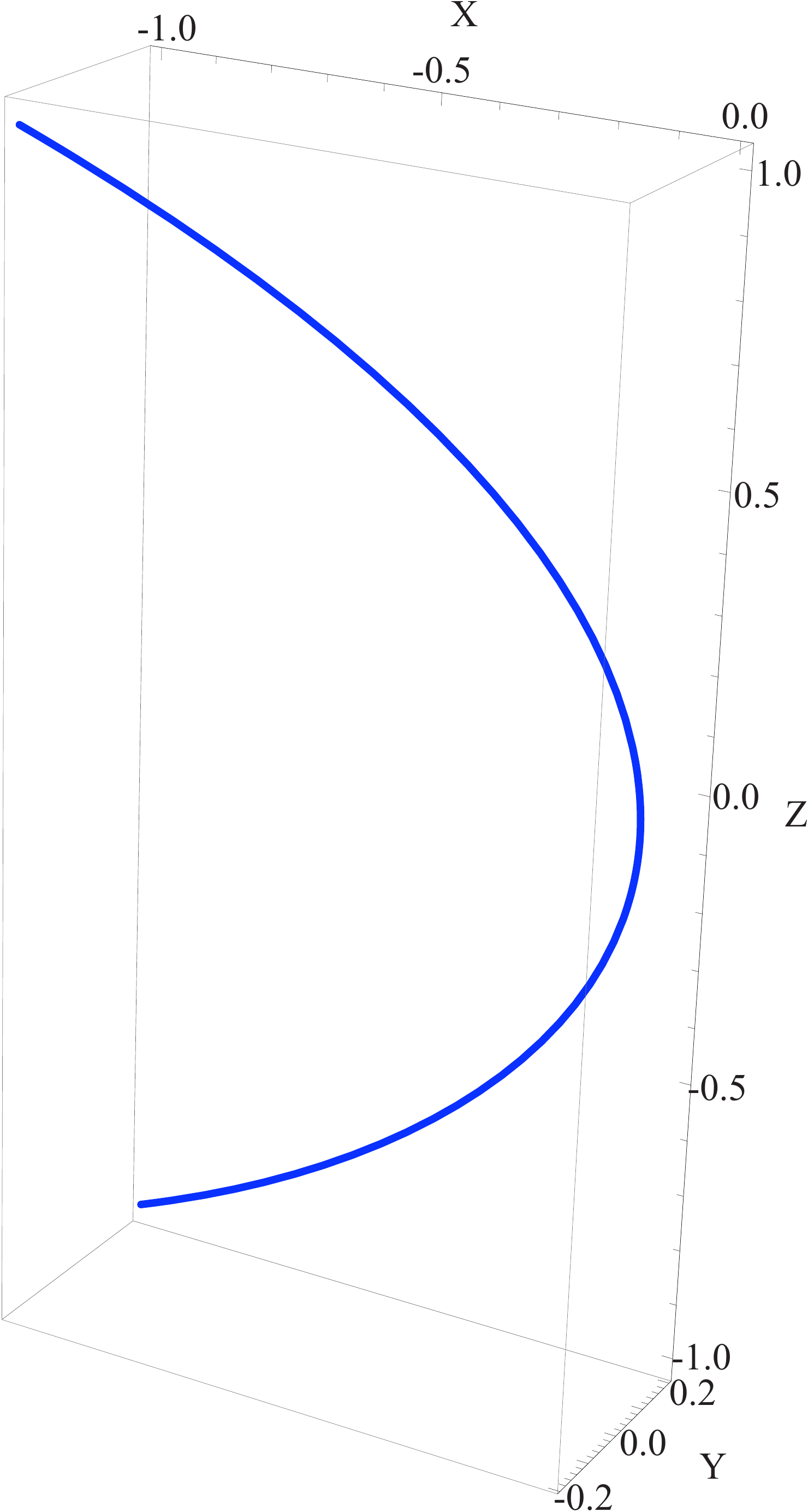}
\caption{Plot of the pseudo-parametrization used, with its distortion through space [Eq.~\eqref{pseudopam} with $a=1$, $k =2$ and $b=5$]}
\label{fig:coords2}
\end{figure}

Furthermore, helices have no variation in $\rho$. Hence for the complete curve to take any form, a constant pseudo-curvature $k$, between $\rho$ and $z$, needs to be added to the pseudo
-parametrization [$k$ being defined as $ \approx \sqrt{\rho''(s)^{2} + z''(s)^{2}}$ ]. Keeping $z(s)$'s form [$z(s) = s$],  $\rho''(s)$  equals $-k$. Hence the pseudo-parametrization becomes [Fig.~\ref{fig:coords2}]
\begin{align}
\mathbf{c}(s) = &\left\{ \left(a-\frac{k}{2} s^{2}\right) \textbf{ } \cos\left(\frac{s}{b}\right) + \alpha s, \right. \nonumber \\
& \mbox{ } \left. \left(a-\frac{k}{2} s^{2}\right)\textbf{ } \sin\left(\frac{s}{b}\right)- \left(\frac{a}{b}  - \beta\right) s, s \right\} , \label{pseudopam}
\end{align} 
where $\alpha$ and $\beta$ are the small angles of deviation from $\mathbf{\hat{z}}$ in the $\mathbf{\hat{x}}$ and $\mathbf{\hat{y}}$ directions respectively, i.e. $\left.\mathbf{\hat{t}}\right|_{s=0} = \left\{\alpha,\beta,1\right\} $.

This parametrization makes the curvature, torsion, and local derivatives in $s$, to first order in $\alpha$ and $\beta$, to be
\begin{gather}
\kappa = \frac{a}{b^{2}} + k, \\
\tau = \frac{a + 3 b^{2} k}{ b( a +  b^{2} k)},\\
\frac{\partial}{\partial s} \rightarrow \frac{\partial}{\partial z} + \alpha \frac{\partial}{\partial x} +\beta \frac{\partial}{\partial y},
\end{gather}
and
\begin{equation} \label{2nd}
\frac{\partial^{2}}{\partial s^{2}} \rightarrow \frac{\partial^{2}}{\partial z^{2}} - \kappa \frac{\partial}{\partial x} + 2 \alpha \frac{\partial^{2}}{\partial x \mbox{ } \partial z} + 2\beta \frac{\partial^{2}}{\partial y \mbox{ } \partial z}.
\end{equation}
This makes the gradient and Laplacian operators ($\nabla'$ and $\nabla'^{2}$)
\begin{equation}\label{a}
\nabla' = \nabla + \left(\alpha \frac{\partial}{\partial x} +\beta \frac{\partial}{\partial y}\right) \mathbf{\hat{z}},
\end{equation}
and
\begin{equation}\label{b}
\nabla'^{2} = \nabla^{2} - \kappa \frac{\partial}{\partial x} + 2 \alpha \frac{\partial^{2}}{\partial x \mbox{ } \partial z} + 2\beta \frac{\partial^{2}}{\partial y \mbox{ } \partial z},
\end{equation}
where $\nabla$ has the standard Cartesian definition ($ \nabla = \{ \partial_{x}, \partial_{y},\partial_{z} \}$).

Using these to transform the Gross-Pitaevskii equation [Eq.~\eqref{gp}] into the local coordinates around a point on the vortex line $\mathbf{r_{0}}$, assuming that $\mathbf{\Omega}$ is in $\mathbf{\hat{z}}$, and assuming that the vortex line has no velocity in $\mathbf{\hat{z}}$, the equation for the behaviour near the vortex core becomes
\begin{gather} 
-\frac{\hbar^{2}}{2 M}\left( \nabla^{2} - \kappa \frac{\partial}{\partial x} + 2 \alpha \frac{\partial^{2}}{\partial x \mbox{ } \partial z} + 2\beta \frac{\partial^{2}}{\partial y \mbox{ } \partial z}\right) \Psi \nonumber \\
+  V_{tr}(\mathbf{r_{0}}) \mbox{ } \Psi   + g |\Psi|^{2} \mbox{ } \Psi - \mu(\Omega) \mbox{ } \Psi   \nonumber \\ 
+ \Psi \mbox{ } \mathbf{r} \cdot \left[\nabla+ \left(\alpha \frac{\partial}{\partial x} +\beta \frac{\partial}{\partial y}\right) \mathbf{\hat{z}}\right] V_{tr}(\mathbf{r_{0}}) \nonumber \\
+ i\hbar (\mathbf{\Omega} \times  \mathbf{r_{0}}) \cdot \nabla \Psi = i \hbar \left(\frac{\partial \Psi}{\partial t} - \mathbf{V}\cdot \nabla \Psi\right) \nonumber \\ \label{A}\end{gather} 
where the trap terms have been expanded to 1st order (order $r_{c}$), $\mathbf{r}$ is the position vector in the local coordinate system, $\mathbf{\Omega}$ is the rotation vector and $\mathbf{V}$ is the velocity of the vortex line at $\mathbf{r_{0}}$.

This equation describes the behaviour of the condensate near the vortex core, in terms of the properties of the vortex core itself. Hence, solving this equation gives the desired inner region behaviour of the BEC.

\section{The behaviour near the vortex core} \label{sec:inner}
The steady state solutions of Eq.~\eqref{A} describes a condensate containing a vortex with standing waves along the line of the vortex. Hence to determine a relation for these standing waves, the inner equation to solve is
\begin{gather}
-\frac{\hbar^{2}}{2 M}\left( \nabla^{2} - \kappa \frac{\partial}{\partial x} + 2 \alpha \frac{\partial^{2}}{\partial x \mbox{ } \partial z} + 2\beta \frac{\partial^{2}}{\partial y \mbox{ } \partial z}\right) \Psi \nonumber \\
+  V_{tr}(\mathbf{r_{0}}) \mbox{ } \Psi   + g |\Psi|^{2} \mbox{ } \Psi - \mu(\Omega) \mbox{ } \Psi   \nonumber \\ 
+ \Psi \mbox{ } \mathbf{r} \cdot \left[\nabla+ \left(\alpha \frac{\partial}{\partial x} +\beta \frac{\partial}{\partial y}\right) \mathbf{\hat{z}}\right] V_{tr}(\mathbf{r_{0}}) \nonumber \\
 = -i \hbar  \mathbf{V'}\cdot \nabla \Psi
\end{gather}
where $\mathbf{V'}$ = $\mathbf{V} + (\mathbf{\Omega} \times  \mathbf{r_{0}})$.

The small nature of the wave perturbations give four small parameters (assumed to be of similar order) to use for a perturbation expansion: $\kappa$, $\alpha$, $\beta$ and $r_{c}$. The 0th order terms, of this expansion describe a condensate with an unperturbed vortex: 
\begin{equation} \label{psi0}
\left[ -\frac{\hbar^{2}}{2 M} \nabla^{2} + V_{tr}(\mathbf{r_{0}}) + g |\Psi_{0}|^{2} - \mu(\Omega)  \right] \Psi_{0} = 0.
\end{equation}
The 1st order terms account for the perturbations on the vortex line: 
\begin{gather}
 -\frac{\hbar^{2}}{2 M} \nabla^{2} \Psi_{1} + V_{tr}(\mathbf{r_{0}}) \Psi_{1} + 2 g |\Psi_{0}|^{2} \Psi_{1}  + g \Psi_{0}^{2} \Psi_{1}^{*} - \mu(\Omega) \Psi_{1} \nonumber\\
  = -\frac{ \hbar^{2} \kappa}{2 M} \frac{\partial \Psi_{0}}{\partial x} + \frac{  \hbar^{2} \alpha}{M} \frac{\partial^{2} \Psi_{0}}{\partial x \mbox{ } \partial z} + \frac{ \hbar^{2} \beta}{M} \frac{\partial^{2} \Psi_{0}}{\partial y \mbox{ } \partial z}  \nonumber\\ 
   - i \hbar  \mathbf{V'}\cdot \nabla \Psi_{0} -\Psi_{0}\textbf{ } \mathbf{r} \cdot \left[\nabla+ \left(\alpha \frac{\partial}{\partial x} +\beta \frac{\partial}{\partial y}\right) \mathbf{\hat{z}}\right] V_{tr}(\mathbf{r_{0}}). \nonumber\\  \label{psi1}
  \end{gather}

Only the far from core behaviour of the inner solution is used for matching. Hence, the large $\rho=\sqrt{x^2+y^2}$ behaviour of Eq.~\eqref{psi1} is required. In this regime $|\Psi_{0}|$ should be cylindrically symmetric, and can be approximated by the Thomas-Fermi solution $|\Psi_{TF}| \approx \sqrt{ [\mu(\Omega) -  V_{tr}(\mathbf{r_{0}})]/g}$. This {\it simplifies} Eq.~\eqref{psi1}  to
\begin{multline}\label{psi1c2}
-\frac{\hbar^{2}}{2 M} \nabla^{2} \Psi_{1} + g |\Psi_{0}|^{2} \Psi_{1} + g \Psi_{0}^{2} \Psi_{1}^{*} \\
 =-\Psi_{0} z \textbf{ } \partial_{z'} V_{tr} (\mathbf{r_{0}}) \\ 
+ \cos(\phi) \left[\frac{\hbar^{2} \alpha}{M} \frac{\partial^{2} \Psi_{0}}{\partial \rho \mbox{ } \partial z} -\Psi_{0}\textbf{ }  \rho \textbf{ } |\nabla V_{tr} (\mathbf{r_{0}})|_{x}   \right.     \\
 \quad +\frac{\hbar^{2} \beta}{ M\rho}\frac{\partial^{2} \Psi_{0}}{\partial \phi \mbox{ } \partial z} -  \frac{i \hbar  V'_{y}}{\rho} \frac{\partial \Psi_{0}}{\partial \phi}   \\
 \mbox{ } \left. - \left(\frac{\hbar^{2}\kappa}{2M} + i \hbar  V'_{x}\right)  \frac{\partial \Psi_{0}}{\partial \rho}     \right] \\
 + \sin(\phi) \left[ \frac{\hbar^{2}\beta}{M}\frac{\partial^{2} \Psi_{0}}{\partial \rho \mbox{ } \partial z}  -\Psi_{0} \textbf{ } \rho \textbf{ } |\nabla V_{tr} (\mathbf{r_{0}})|_{y} \right.  \\
  \mbox{ } - \frac{ \hbar^{2} \alpha}{ M \rho} \frac{\partial^{2} \Psi_{0}}{\partial \phi \mbox{ } \partial z} - i \hbar  V'_{y} \frac{\partial \Psi_{0}}{\partial \rho}   \\
 \mbox{ } \left. + \frac{1}{\rho}\left(\frac{\hbar^{2}\kappa}{2M} + i \hbar  V'_{x}\right) \frac{\partial \Psi_{0}}{\partial \phi}  \right], 
\end{multline}
where $\partial_{z'} = \partial_{z} + \alpha \partial_{x} + \beta \partial_{y}$, and $|\nabla V_{tr} (\mathbf{r_{0}})|_{i}$ and $V'_{i}$ are the trap gradient and vortex line velocity in the $i$ direction.

Assuming that the wave function has the form
\begin{align}
\Psi =& \left[ \left| \Psi_{0} \right| + D(\rho,z) +\chi(\rho,z)\cos(\phi)\right.\nonumber \\ 
&\mbox{ }+ \left.\zeta(\rho,z)\sin(\phi)\right] e^{i q \phi + i \eta(\rho,z) \cos(\phi) + i \lambda(\rho,z) \sin(\phi)}, \nonumber \\
\label{psi}
\end{align} 
makes $ \Psi_{0} = \left| \Psi_{0} \right|  e^{i q \phi}$ and 
\begin{align}
\Psi_{1} =& \left[D(\rho,z)+ \chi(\rho,z) \cos(\phi) \right. \nonumber \\
 &\mbox{ }+\zeta(\rho,z) \sin(\phi) + i \left| \Psi_{0} \right| \eta(\rho,z) \cos(\phi) \nonumber \\ 
 &\quad+ \left. i \left| \Psi_{0} \right| \lambda(\rho,z) \sin(\phi)\right] e^{i q \phi}, \label{form} 
\end{align} 
where $q$ is the winding number of the vortex.

This reduces Eq.~\eqref{psi1c2} into five coupled differential equations describing the behaviour of the perturbation functions $D$, $\chi$, $\zeta$, $\eta$, and $\lambda$:

\begin{multline}
\frac{1}{\rho} \frac{\partial}{\partial \rho} (\rho \frac{\partial D}{\partial \rho} ) + \frac{\partial^{2} D}{\partial z^{2}} - \frac{q^{2} D}{\rho^{2}} - \frac{4 M g \left| \Psi_{0} \right|^{2} D}{\hbar^{2}} \\
 = \frac{ 2 M |\Psi_{0}| \textbf{ }  z}{\hbar^{2}} \textbf{ } \partial_{z'} V_{tr} (\mathbf{r_{0}}) ,  \label{D}
\end{multline}

\begin{multline}
\frac{1}{\rho} \frac{\partial}{\partial \rho} (\rho \frac{\partial\chi}{\partial \rho} ) + \frac{\partial^{2} \chi}{\partial z^{2}} - \frac{(q^{2} +1)\chi}{\rho^{2}} - \frac{2q\left| \Psi_{0} \right| \lambda}{\rho^{2}}   \\
 - \frac{4 M g \left| \Psi_{0} \right|^{2} \chi}{\hbar^{2}} = \frac{2 M  \left| \Psi_{0} \right|  \left| \nabla V_{tr} \right|_{x} \rho}{\hbar^{2}} \\
- \frac{2 M q \left| \Psi_{0} \right| V'_{y}}{\hbar \rho} + \kappa \frac{ \partial \left| \Psi_{0} \right|}{ \partial \rho} - 2 \alpha \frac {\partial^{2} \left| \Psi_{0} \right|}{\partial \rho \mbox{ } \partial z},  \label{chi} 
\end{multline}

\begin{multline} \label{zeta}
\frac{1}{\rho} \frac{\partial}{\partial \rho} (\rho \frac{\partial\zeta}{\partial \rho} ) + \frac{\partial^{2} \zeta}{\partial z^{2}} - \frac{(q^{2} +1)\zeta}{\rho^{2}} + \frac{2q\left| \Psi_{0} \right| \eta}{\rho^{2}}  \\
- \frac{4 M g \left| \Psi_{0} \right|^{2} \zeta}{\hbar^{2}} =  \frac{2 M  \left| \Psi_{0} \right|  \left| \nabla V_{tr} \right|_{y} \rho}{\hbar^{2}} \\   + \frac{2 M q \left| \Psi_{0} \right| V'_{x}}{\hbar \rho} - 2 \beta \frac {\partial^{2} \left| \Psi_{0} \right|}{\partial \rho \mbox{ } \partial z},
\end{multline}

\begin{multline} \label{eta}
\left| \Psi_{0} \right| \left(\frac{\partial^{2}}{\partial \rho^{2}} + \frac{1}{\rho}\frac{\partial}{\partial \rho}  + \frac{\partial^{2} }{\partial z^{2}} - \frac{ 1}{\rho^{2}}\right) \eta \\
 + 2 \left( \frac{\partial \left| \Psi_{0} \right|}{\partial \rho} \frac{\partial \eta}{\partial \rho} +  \frac{\partial \left| \Psi_{0} \right|}{\partial z} \frac{\partial \eta}{\partial z} + \frac{ q \zeta}{\rho^{2}} \right)   \\
=\frac{ 2 M V'_{x}}{\hbar}  \frac{\partial \left| \Psi_{0} \right|}{\partial \rho} - \frac{ 2 q \beta}{ \rho} \frac{\partial \left| \Psi_{0} \right|}{\partial z},
\end{multline}

\begin{multline} \label{lambda}
\left| \Psi_{0} \right| \left(\frac{\partial^{2}}{\partial \rho^{2}} + \frac{1}{\rho}\frac{\partial}{\partial \rho}  + \frac{\partial^{2} }{\partial z^{2}} - \frac{ 1}{\rho^{2}}\right) \lambda  \\
 + 2 \left( \frac{\partial \left| \Psi_{0} \right|}{\partial \rho} \frac{\partial \lambda}{\partial \rho} +  \frac{\partial \left| \Psi_{0} \right|}{\partial z} \frac{\partial \lambda}{\partial z} - \frac{ q \chi}{\rho^{2}} \right)  \\
= \frac{ 2 M V'_{y}}{\hbar}  \frac{\partial \left| \Psi_{0} \right|}{\partial \rho} + \frac{ 2 q \alpha}{ \rho} \frac{\partial \left| \Psi_{0} \right|}{\partial z} - \frac{ q \kappa \left|\Psi_{0}\right|}{ \rho}.
\end{multline}

These equations separate into two connected groups: equations describing perturbations on condensate density [Eqs.~\eqref{D}, \eqref{chi}, and \eqref{zeta}], and equations describing perturbations on the condensate's phase [Eqs.~\eqref{eta} and \eqref{lambda}]. Different groups having different assumptions and transformations that simplify the set of equations.
\begin{itemize}
\item Substituting $\eta$ and $\lambda$ with $\overline{\eta}+  \rho V'_{x} M/\hbar$ and $\overline{\lambda}+  \rho V'_{x} M/\hbar$, removes $\mathbf{V'}$ dependence from the describing equations. As $\rho$ is a solution to $\partial_{\rho}^{2} f(\rho) + \partial_{\rho} f(\rho)/\rho  -  f(\rho)/\rho^{2} = 0$, an  arbitrary constant multiplied by $\rho$ can be added to $f(\rho)$ without changing its final solution; consequently, this constant should be chosen to best simplify the equations, removing the $\mathbf{V'}$ dependence.
\item Derivatives of the phase terms with respect to $z$ can be omitted (and by association $\overline{\eta}$ and $\overline{\lambda}$). The gradient of the phase defines the condensates velocity, variation in $z$ therefore inducing flow in $\mathbf{\hat{z}}$. This kind of flow is caused by either the addition of sources and sinks or vortex rings to the BEC. This derivation only considers the behaviour of a `straight' vortex in a condensate, and therefore should have no velocity in $\mathbf{\hat{z}}$.
\item The phase terms $\partial_{\rho} |\Psi_{0}|  \partial_{\rho} \overline{\eta}$ and $\partial_{\rho} |\Psi_{0}| \partial_{\rho} \overline{\lambda}$ can be removed. In Eqs.~\eqref{eta} and \eqref{lambda} there are two terms involving the 1st derivatives of the phase with respect to $\rho$, having coefficients $|\psi_{0}|/\rho$ and $\partial_{\rho} |\Psi_{0}|$.  As mentioned previously, at large $\rho$, $|\Psi_{0}|$ is approximately $|\Psi_{TF}|$, and varies on length scales of the order trap radius $R_{\perp}$; therefore the change in $|\Psi_{0}|$ with respect to $\rho$ (varies on $r_{c}$) is small compared with $|\psi_{0}|/\rho$, which is approximately constant, and so has little effect.
\item Kinetic energy terms, from the waves density, can be ignored.  In the Thomas-Fermi regime the kinetic energy from the waves density is assumed negligible, giving the well known inverted parabola structure ($g | \Psi_{TF}|^{2} = \mu_{TF} - V_{tr}$). As this behaviour is expected in the matching region, the Laplacians ($\nabla^{2}$) of Eqs.~\eqref{D}, \eqref{chi} and \eqref{zeta} are approximately 0, removing the differential behaviour from these equations.
\end{itemize}
Applying these assumptions and rearranging Eqs.~\eqref{D}, \eqref{chi}, \eqref{zeta}, \eqref{eta}, and \eqref{lambda}, we find

\begin{align}
  D(\rho , z) =& - \frac{ \textbf{ }  z}{2 g \left| \Psi_{0} \right|} \textbf{ } \partial_{z'} V_{tr} (\mathbf{r_{0}}) , \label{d2} \\ \nonumber \\
 \chi(\rho , z) =& -\frac{  \left| \nabla V_{tr} \right|_{x} \rho}{2 g \left| \Psi_{0} \right|}  - \frac{\hbar^{2} \kappa}{ 4 M g \left| \Psi_{0} \right|^{2}} \frac{ \partial \left| \Psi_{0} \right|}{ \partial \rho}\nonumber \\
  &+ \frac{ \hbar^{2} \alpha}{2 M g \left| \Psi_{0} \right|^{2}} \frac {\partial^{2} \left| \Psi_{0} \right|}{\partial \rho \mbox{ } \partial z}  - \frac{ q \hbar^{2}  \overline{\lambda}}{2 M g \left| \Psi_{0} \right|\rho^{2}}, \label{chi2} \\ \nonumber \\
  \zeta(\rho , z) =& - \frac{\left| \nabla V_{tr} \right|_{y} \rho}{2 g \left| \Psi_{0} \right|}  + \frac{ \hbar^{2} \beta}{ 2 M g\left| \Psi_{0} \right|^{2}} \frac {\partial^{2} \left| \Psi_{0} \right|}{\partial \rho \mbox{ } \partial z} \nonumber \\ 
  &+ \frac{ q \hbar^{2} \overline{\eta}}{2 M g \left| \Psi_{0} \right|\rho^{2}}, \label{zeta2}
 \end{align}
\begin{gather}
\left| \Psi_{0} \right| \left(\frac{\partial^{2}}{\partial \rho^{2}} + \frac{1}{\rho}\frac{\partial}{\partial \rho} - \frac{ 1}{\rho^{2}}\right) \overline{\eta}  + \frac{ 2 q \zeta}{\rho^{2}} \nonumber \\
 = - \frac{ 2 q \beta}{ \rho} \frac{\partial \left| \Psi_{0} \right|}{\partial z}, \label{eta2} \\ \nonumber \\
\left| \Psi_{0} \right| \left(\frac{\partial^{2}}{\partial \rho^{2}} + \frac{1}{\rho}\frac{\partial}{\partial \rho}  - \frac{ 1}{\rho^{2}}\right) \overline{\lambda} - \frac{ 2 q \chi}{\rho^{2}} \nonumber \\
= \frac{ 2 q \alpha}{ \rho} \frac{\partial \left| \Psi_{0} \right|}{\partial z} - \frac{ q \kappa \left|\Psi_{0}\right|}{ \rho}. \label{lambda2}
\end{gather}

Equation \eqref{d2} solves for the perturbation $D(\rho , z)$, being the first order correction to the local Thomas-Fermi profile, i.e. correcting for the changing condensate density in the z direction.

The remaining density perturbations, Eqs.~\eqref{chi2} and \eqref{zeta2}, are linear. These equations contain appropriate Thomas-Fermi, curvature and $\mathbf{\hat{z}}$ deviation correction terms and a term that depends on a phase perturbation. As a consequence, these will be solved once the phase perturbations are known. Using these to remove $\chi$ and $\zeta$  dependence from Eqs.~\eqref{eta2} and \eqref{lambda2}, two independent differential equations for $\overline{\eta}$ and $\overline{\lambda}$ are found:
\begin{multline}\label{1a}
\left(\frac{\partial^{2}}{\partial \rho^{2}} + \frac{1}{\rho}\frac{\partial}{\partial \rho} - \frac{ 1}{\rho^{2}}\left(1-\frac{ q^{2} \hbar^{2}}{ M g \left| \Psi_{0} \right|\rho^{2}}\right) \right) \overline{\eta} \\ 
=-\frac{ 2  q \beta}{ \left| \Psi_{0} \right|  \rho} \frac{\partial \left| \Psi_{0} \right|}{\partial z} + \frac{ q \left| \nabla V_{tr} \right|_{y}}{g \left| \Psi_{0} \right|^{2} \rho} \\ 
 - \frac{ \hbar^{2} q \beta}{M g \left| \Psi_{0} \right|^{3} \rho^{2}} \frac {\partial^{2} \left| \Psi_{0} \right|}{\partial \rho \mbox{ } \partial z} , 
\end{multline}
\begin{multline}
\left(\frac{\partial^{2}}{\partial \rho^{2}} + \frac{1}{\rho}\frac{\partial}{\partial \rho}  - \frac{ 1}{\rho^{2}}\left(1- \frac{  \hbar^{2} q^{2}}{ M g \left| \Psi_{0} \right|\rho^{2}} \right)\right) \overline{\lambda}  \\
=\frac{ 2 q \alpha}{ \left| \Psi_{0} \right| \rho} \frac{\partial \left| \Psi_{0} \right|}{\partial z}- \frac{  q \left| \nabla V_{tr} \right|_{x}}{g \left| \Psi_{0} \right|^{2} \rho} - \frac{ \hbar^{2} q \kappa}{2 M  g \left| \Psi_{0} \right|^{3} \rho^{2}} \frac{ \partial \left| \Psi_{0} \right|}{ \partial \rho} \\
+ \frac{ \hbar^{2} q \alpha}{M g \left| \Psi_{0} \right|^{3} \rho^{2}} \frac {\partial^{2} \left| \Psi_{0} \right|}{\partial \rho \mbox{ } \partial z}  - \frac{ q \kappa}{ \rho} . \label{2a}
\end{multline}

Comparing the relative terms,  $\frac{  \hbar^{2} q^{2}}{ M g \left| \Psi_{0} \right|\rho^{2}}$, $\frac{ \hbar^{2} q \kappa}{2 M  g \left| \Psi_{0} \right|^{3} \rho^{2}} \frac{ \partial \left| \Psi_{0} \right|}{ \partial \rho}$, $ \frac{ \hbar^{2} q \alpha}{M g \left| \Psi_{0} \right|^{3} \rho^{2}} \frac {\partial^{2} \left| \Psi_{0} \right|}{\partial \rho \mbox{ } \partial z}$, and $\frac{ \hbar^{2} q \beta}{M g \left| \Psi_{0} \right|^{3} \rho^{2}} \frac {\partial^{2} \left| \Psi_{0} \right|}{\partial \rho \mbox{ } \partial z}$ are much smaller than their corresponding counterparts by at least an order of $1/R_{\perp}$ (especially as $\rho$ gets large). Consequently, these terms do not significantly affect the far from core behaviour and can be omitted, turning Eqs.~\eqref{1a} and \eqref{2a} into

\begin{align}
\left(\frac{\partial^{2}}{\partial \rho^{2}} + \frac{1}{\rho}\frac{\partial}{\partial \rho} - \frac{ 1}{\rho^{2}} \right) \overline{\eta}  \approx & \frac{ q \beta}{  g \left| \Psi_{TF} \right|^{2}  \rho} \frac{\partial V_{tr}}{\partial z} \nonumber \\ 
&+ \frac{ 2 q \left| \nabla V_{tr} \right|_{y}}{g \left| \Psi_{TF} \right|^{2} \rho} , \label{1} \\
\left(\frac{\partial^{2}}{\partial \rho^{2}} + \frac{1}{\rho}\frac{\partial}{\partial \rho}  - \frac{ 1}{\rho^{2}}\right) \overline{\lambda} \approx & - \frac{  q \alpha}{ g \left| \Psi_{TF} \right|^{2} \rho} \frac{\partial V_{tr}}{\partial z} - \frac{q \kappa}{\rho} \nonumber \\
 &- \frac{  q \left| \nabla V_{tr} \right|_{x}}{g \left| \Psi_{TF} \right|^{2} \rho} , \label{2}
\end{align} 
where $|\Psi_{0}|$ has been replaced with $|\Psi_{TF}|$.

Given that $|\Psi_{TF}|$, $\left| \nabla V_{tr} \right|$ and $\frac{\partial V_{tr}}{\partial z}$ are constant in the inner solution expansion, these equations are Euler homogeneous equations, and have solutions of the form $\rho^{n}$. Solving these equations, and making them non-divergent at $\rho = 0$ the solutions become

\begin{align}
\eta =& \left( \frac{q \left| \nabla V_{tr} \right|_{y}}{2 g \left|\Psi_{TF}\right|^{2}} + \frac{ q \beta}{2 g \left|\Psi_{TF}\right|^{2}} \frac{\partial V_{tr}}{\partial z} \right) \rho \ln(\rho)  \nonumber \\ 
&+ \frac{M}{\hbar} \rho V'_{x}+ \rho \mbox{ } A, \label{etaf} \\
\lambda =& -\left( \frac{q \left| \nabla V_{tr} \right|_{x}}{2 g \left|\Psi_{TF}\right|^{2}} + \frac{ q \alpha}{2 g \left|\Psi_{TF}\right|^{2}} \frac{\partial V_{tr}}{\partial z} + \frac{q \kappa }{2} \right) \rho \ln(\rho) \nonumber \\ 
&+ \frac{M}{\hbar} \rho V'_{y} + \rho \mbox{ } B, \label{lambdaf}
\end{align}
where $A$ and $B$ are unknown constants that can depend on $|\Psi_{TF}|$, $\left| \nabla V_{tr} \right|$, $\kappa$ and $\frac{\partial V_{tr}}{\partial z}$.

Therefore the phase of the inner solution is
\begin{align}
S =& \quad q \phi + \left( \frac{q \left| \nabla V_{tr} \right|_{y}}{2 g \left|\Psi_{TF}\right|^{2}} + \frac{ q \beta}{2 g \left|\Psi_{TF}\right|^{2}} \frac{\partial V_{tr}}{\partial z} \right) x  \ln( \rho)  \nonumber \\
 &- \left( \frac{q \left| \nabla V_{tr} \right|_{x}}{2 g \left|\Psi_{TF}\right|^{2}} + \frac{ q \alpha}{2 g \left|\Psi_{TF}\right|^{2}} \frac{\partial V_{tr}}{\partial z} + \frac{q \kappa }{2} \right) y \ln( \rho)  \nonumber \\
&+ x \mbox{ } A +y \mbox{ } B + \frac{M}{\hbar} x V'_{x} + \frac{M}{\hbar} y V'_{y}
\end{align}
or
\begin{align} 
S =&\quad  q \phi +\mathbf{r} \cdot \frac{M}{\hbar} \left( \mathbf{V}+ \mathbf{\Omega} \times \mathbf{r_{0}}\right) \nonumber \\ 
 &-\mathbf{r} \cdot \frac{q \ln(\rho)}{2 g \left|\Psi_{TF}\right|^{2} } \left[ \kappa g \left|\Psi_{TF}\right|^{2} \mathbf{\hat{b}} + \mathbf{\hat{z}} \times \nabla V_{tr}(\mathbf{r_{0}})\right]\nonumber \\
  &+ \mathbf{r} \cdot \frac{q \ln(\rho)}{2 g \left|\Psi_{TF}\right|^{2} } \left[ \left(\mathbf{\hat{z}} \times \mathbf{\hat{t}}\right)\left.\frac{\partial V_{tr}}{\partial z}\right|_{\mathbf{r}=\mathbf{r_{0}}} \right] \nonumber \\ 
&+ \mathbf{r} \cdot \mathbf{E}\left[\kappa \mathbf{\hat{b}}, \frac{\mathbf{\hat{z}} \times \nabla V_{tr}(\mathbf{r_{0}})}{g \left|\Psi_{TF}\right|^{2}},\frac{\mathbf{\hat{z}} \times \mathbf{\hat{t}}}{g \left|\Psi_{TF}\right|^{2}}\left. \frac{\partial V_{tr}}{\partial z}\right|_{\mathbf{r}=\mathbf{r_{0}}}, \mathbf{r_{0}} \right] , \nonumber \\ \label{inner}
\end{align}
as $\mathbf{\hat{t}} = \lbrace \alpha, \beta, 1 \rbrace$, and $\mathbf{E} = \{A, B , 0\}$. The dependant terms in $\mathbf{E}$ are written down here to emphasise that the constant can depend on $|\Psi_{TF}|$, $\left| \nabla V_{tr} \right|$, $\kappa$ and $\frac{\partial V_{tr}}{\partial z}$, with the structure of the dependants chosen because of the structure of the other terms in Eq.~\eqref{inner}.

 Using Eqs.~\eqref{etaf} and \eqref{lambdaf}, the density perturbations are

\begin{align}
  D =& - \frac{ \textbf{ }  z}{2 g \left| \Psi_{TF} \right|} \textbf{ } \partial_{z'} V_{tr} (\mathbf{r_{0}}) ,\label{df} \\ \nonumber \\
 \chi =& -\frac{  \left| \nabla V_{tr} \right|_{x} \rho}{2 g \left| \Psi_{TF} \right|}  - \frac{\hbar^{2} \kappa}{ 4 M g \left| \Psi_{TF} \right|^{2}} \frac{ \partial \left| \Psi_{TF} \right|}{ \partial \rho} \nonumber \\
 &+\frac{\hbar^{2} q^{2} \left( \left| \nabla V_{tr} \right|_{x} +  \alpha \frac{\partial V_{tr}}{\partial z} +  \kappa g \left| \Psi_{TF} \right|^{2}  \right) \ln(\rho)}{4 M g^{2} \left| \Psi_{TF} \right|^{3}\rho} \nonumber \\
 &+  \frac{ \hbar^{2} \alpha}{2 M g \left| \Psi_{TF} \right|^{2}} \frac {\partial^{2} \left| \Psi_{TF} \right|}{\partial \rho \mbox{ } \partial z} -\frac{q \hbar^{2} B}{2 M g \left| \Psi_{TF} \right| \rho}  ,  \label{chif} \\ \nonumber \\
  \zeta =& - \frac{\left| \nabla V_{tr} \right|_{y} \rho}{2 g \left| \Psi_{TF} \right|}  + \frac{ \hbar^{2} \beta}{ 2 M g\left| \Psi_{TF} \right|^{2}} \frac {\partial^{2} \left| \Psi_{TF} \right|}{\partial \rho \mbox{ } \partial z}  \nonumber \\
  &+ \frac{ q^{2} \hbar^{2} \left(  \left| \nabla V_{tr} \right|_{y} +  \beta \frac{\partial V_{tr}}{\partial z} \right)  \ln(\rho)}{4 M g^{2} \left| \Psi_{TF} \right|^{3} \rho} \nonumber \\
   &+ \frac{q \hbar^{2} A}{2 M g \left| \Psi_{TF} \right| \rho}. \label{zetaf}
\end{align}
These density perturbations are independent of the vortex core's velocity ($\mathbf{V}$), being, to lowest order, corrections to the local Thomas-Fermi profile. Hence these solutions already match the Thomas-Fermi density profile of the outer region. This indicates that the perturbations on a vortex line predominantly affects the inner phase of the condensate [Eq.~\eqref{inner}], which through matching to the outer regions phase, should give the desired vortex relation.
 
\section{The phase of the outer region} \label{sec:outer}

Far from the vortex core, the BEC's wave function density has the Thomas-Fermi profile, and its phase ($S$) solves
\begin{gather} 
\nabla \cdot \left(|\Psi_{TF}|^{2} \nabla S\right) - \frac{M}{\hbar} \mathbf{\Omega}\cdot\left(\mathbf{r} \times \nabla \right) |\Psi_{TF}|^{2} = 0, \label{GPO} \\ \nonumber \\
\mathbf{\hat{z}}\cdot \nabla \times \left( \nabla S \right) = 2 \pi q \delta^{(2)} (\mathbf{\rho}-\mathbf{\rho_{0}}),\label{V}
\end{gather}
where $\delta^{(2)} (\mathbf{\rho}-\mathbf{\rho_{0}})$ is the delta function in $\rho$ and $\phi$.

These equations are linear, and therefore allows $S$ to be solved in two separate components: one dealing with rotation [the particular solution of Eq.~\eqref{GPO}, $S_{\Omega}$] and one with everything else ($S_{c}$):
\begin{gather} 
 \nabla\cdot \left( |\Psi_{TF}|^{2} \nabla S_{\Omega} \right) - \frac{M}{\hbar} \mathbf{\Omega}\cdot\left(\mathbf{r} \times \nabla \right) |\Psi_{TF}|^{2} = 0,  \label{SoGPO} \\ \nonumber \\
 \nabla \times \left( \nabla S_{\Omega} \right) = 0, \label{SoV} \\ \nonumber \\
 \nabla \cdot \left( |\Psi_{TF}|^{2} \nabla S_{c} \right) = 0, \label{ScGPO} \\ \nonumber \\
\mathbf{\hat{z}}\cdot \nabla \times \left( \nabla S_{c} \right) =  2 \pi q \delta^{(2)} (\mathbf{\rho}-\mathbf{\rho_{0}}).  \label{ScV}
\end{gather}

Assuming $\mathbf{\Omega} = \Omega \mathbf{\hat{z}}$ Eqs.~\eqref{SoGPO} and \eqref{SoV} have the solution
\begin{equation} \label{f}
S_{\Omega} = - \frac{M}{\hbar} \frac{(\omega_{x}^{2} - \omega_{y}^{2})}{(\omega_{x}^{2} + \omega_{y}^{2})} \Omega x y ,
\end{equation}
where $\omega_{x}$ and $\omega_{y}$ are the trapping frequencies in $x$ and $y$, respectively.

Transforming into the local coordinate system of the line ($\mathbf{r} \rightarrow \mathbf{r} + \mathbf{r_{0}}$), Eq.~\eqref{f} becomes
\begin{equation}\label{So}
S_{\Omega}  \rightarrow \frac{M}{\hbar} \left( ( \mathbf{\Omega} \times \mathbf{r_{0}} ) + 2 \frac{ \nabla V_{tr} (\mathbf{r_{0}}) \times \mathbf{\Omega} }{\nabla_{\perp}^{2} V_{tr}(\mathbf{r_{0}})} \right) \cdot \mathbf{r}.
\end{equation}

$S_{c}$ is not as simple to solve, with no exact solution existing for arbitrary $\rho_{0}$. The gradient of $S_{c}$, however, can be determined and consequently compared with its inner solution equivalent. To do this $\nabla S_{c}$ is written as $- q (\nabla \times (|\Psi_{TF}|^{2} \Phi \mathbf{\hat{z}}))/|\Psi_{TF}|^{2}$, so that it automatically satisfies Eq.~\eqref{ScGPO}, and turns Eq.~\eqref{ScV} into
\begin{eqnarray}
\mathbf{\hat{z}}\cdot \nabla \times \left( \nabla S_{c} \right) &=& \nabla\cdot (\nabla S_{c} \times \mathbf{\hat{z}}) \nonumber \\
&=& \nabla\cdot \left(\nabla_{\perp} \Phi + \Phi \nabla_{\perp} \ln(|\Psi_{TF}|^{2}) \right) \nonumber \\
&=& 2 \pi \delta^{(2)} (\mathbf{\rho}-\mathbf{\rho_{0}}),
\end{eqnarray}
which through rearrangement becomes
\begin{gather}
e^{-\ln|\Psi_{TF}|} \nabla_{\perp}^{2}\left(\Phi e^{\ln|\Psi_{TF}|}\right) -\nabla_{\perp}^{2}\left( e^{-\ln|\Psi_{TF}|}\right) \Phi e^{\ln|\Psi_{TF}|}  \nonumber \\
 = 2 \pi \delta^{(2)} (\mathbf{\rho}-\mathbf{\rho_{0}}). \label{q}
\end{gather}

Transforming Eq.~\eqref{q} into the local coordinates of the vortex line, and taking $\nabla_{\perp}^{2} e^{-\ln|\Psi_{TF}|}$ to be approximately $e^{-\ln|\Psi_{TF}|}  \nabla_{\perp}^{2} V_{tr}/2 g |\Psi_{TF}|^{2} $, it becomes

\begin{gather} 
 \nabla_{\perp}^{2}\left(\Phi e^{\ln|\Psi_{TF}|}\right) -\frac{ \nabla_{\perp}^{2} V_{tr} }{2 g |\Psi_{TF}|^{2}} \Phi e^{\ln|\Psi_{TF}|} \nonumber \\
  = 2 \pi \delta^{(2)} (\mathbf{\rho}) e^{\ln|\Psi_{TF}|}. \label{3}
\end{gather}

In this outer region there should be no flow in the $\mathbf{\hat{\rho}}$ or $\mathbf{\hat{z}}$, and therefore $\Phi$ is purely a function of $\rho$. Furthermore, J. R. Anglin \cite{PhysRevA.65.063611} showed, in a complete hydrodynamic calculation, flow only arises from the vortex's vorticity, thereby indicating that the homogeneous solution to Eq.~\eqref{3} is irrelevant. Hence the solution to Eq.~\eqref{3} is
\begin{equation}
\Phi = - K_{0} \left( \rho \sqrt{\frac{ \nabla_{\perp}^{2} V_{tr} }{2 g |\Psi_{TF}|^{2}}} \right),
\end{equation}
where $K_{0}$ is a modified Bessel function of the second kind and $V_{tr}$ and $ g |\Psi_{TF}|^{2}$ are implicitly functions of $\mathbf{r_{0}}$.

For small $\rho$,
\begin{equation} \label{Sc}
\Phi \approx \ln\left(\frac{e^{c} \rho}{2 R_{\perp} }\right),
\end{equation}
where $c = 0.577...$ is the Euler constant, and $ \nabla_{\perp}^{2} V_{tr}/2 g |\Psi_{TF}|^{2}$ has been approximated by $ 1/R_{\perp}$ (as $ \nabla_{\perp}^{2}V_{tr} \approx M \omega_{\perp}^{2}$ and $ g |\Psi_{TF}|^{2} \approx \mu_{TF} = \frac{1}{2} M \omega_{\perp}^{2} R_{\perp}^{2}$).

Using this 
\begin{align}
\nabla S_{c} =&\frac{- q} {|\Psi_{TF}|^{2}} \left[\nabla \times |\Psi_{TF}|^{2}  \ln\left(\frac{e^{c} \rho}{2 R_{\perp} }\right) \mathbf{\hat{z}}\right] \nonumber \\
= &- q \nabla \times \ln\left(\frac{e^{c} \rho}{2 R_{\perp} }\right) \mathbf{\hat{z}} \nonumber \\
&- q \ln\left(\frac{e^{c} \rho}{2 R_{\perp} }\right) \nabla \times \ln|\Psi_{TF}|^{2} \mathbf{\hat{z}} \nonumber \\
=& \frac{q}{\rho} \mathbf{\hat{\phi}} - \frac{ q \ln\left(\frac{e^{c} \rho}{2 R_{\perp} }\right)}{ g |\Psi_{TF}|^{2}} \mathbf{\hat{z}} \times \nabla V_{tr}(\mathbf{r_{0}}),  \label{gsco}
\end{align}
having used the identity $\nabla \times A(\mathbf{r}) \mathbf{\hat{z}}  = - \mathbf{\hat{z}} \times \nabla A(\mathbf{r})$.

Therefore, the outer phase is fully described by the Eqs.~\eqref{gsco} and \eqref{So}. Through matching these equations with the inner phase [Eq.~\eqref{inner}] a relationship for the velocity of the vortex line is determined.

\section{Matching the inner and outer solutions} \label{sec:match}

To match the outer and inner solutions, the inner phase must be divided into the same $S_{\Omega}$ and $S_{c}$ structure as the outer solution. Assuming that the core velocity $\mathbf{V}$ can be written as a rotational $\mathbf{V_{\Omega}}$ and other $\mathbf{V_{c}}$ component, Eq.~\eqref{inner} separates into
\begin{equation} \label{Soi}
S_{\Omega_{i}} = \frac{M}{\hbar} \left(\mathbf{ V_{\Omega}} + \mathbf{\Omega} \times \mathbf{r_{0}} \right)\cdot \mathbf{r},
\end{equation}
and
\begin{align}
S_{c_{i}} =& \quad q \phi +\mathbf{r} \cdot \frac{M}{\hbar}  \mathbf{V_{c}} - \mathbf{r} \cdot \frac{q \ln(\rho)}{2} \kappa  \mathbf{\hat{b}}  \nonumber \\ 
&- \mathbf{r} \cdot \frac{q \ln(\rho)}{2 g \left|\Psi_{TF}\right|^{2} } \left[  \mathbf{\hat{z}} \times \nabla V_{tr}(\mathbf{r_{0}}) +\left(\mathbf{\hat{z}} \times \mathbf{\hat{t}}\right)\left.\frac{\partial V_{tr}}{\partial z}\right|_{\mathbf{r}=\mathbf{r_{0}}} \right] \nonumber \\
&+ \mathbf{r} \cdot \mathbf{E}\left[\kappa \mathbf{\hat{b}}, \frac{\mathbf{\hat{z}} \times \nabla V_{tr}(\mathbf{r_{0}})}{g \left|\Psi_{TF}\right|^{2}},\frac{\mathbf{\hat{z}} \times \mathbf{\hat{t}}}{g \left|\Psi_{TF}\right|^{2}}\left. \frac{\partial V_{tr}}{\partial z}\right|_{\mathbf{r}=\mathbf{r_{0}}}, \mathbf{r_{0}} \right] , \nonumber \\ \label{Sci}
\end{align}
which can be compared to the outer $S_{\Omega}$ [Eq.~\eqref{So}] and $S_{c}$ [Eq.~\eqref{gsco}] to find the matching condition for the velocity of the line. 

Comparing Eq.~\eqref{Soi} with Eq.~\eqref{So}, shows that $\mathbf{V_{\Omega}}$  is 
\begin{equation} \label{vo}
\mathbf{V_{\Omega}}(\mathbf{r_{0}}) = 2 \frac{\nabla V_{tr} (\mathbf{r_{0}}) \times \mathbf{\Omega} }{\nabla_{\perp}^{2} V_{tr}(\mathbf{r_{0}})}.
\end{equation}

To determine $\mathbf{V_{c}}$, however, $\nabla S_{c}$ of the two regions must be matched. Using Eq.~\eqref{Sci}, $\nabla S_{c_{i}}$ for large $\rho$ is
\begin{align}
\nabla S_{c_{i}} =& \quad \frac{q}{\rho}\mathbf{\hat{\phi}} + \frac{M}{\hbar}  \mathbf{V_{c}}(\mathbf{r_{0}}) -  \frac{q \mbox{ } \kappa  \mathbf{\hat{b}}   }{2}  \cdot \left[\ln(\rho) +\mathbf{\hat{\rho}} \otimes \mathbf{\hat{\rho}} \right] \nonumber \\
&-  \frac{q\mbox{ } \mathbf{\hat{z}} \times \nabla V_{tr}(\mathbf{r_{0}})}{2 g \left|\Psi_{TF}\right|^{2} }  \cdot \left[\ln(\rho) +\mathbf{\hat{\rho}} \otimes \mathbf{\hat{\rho}} \right]    \nonumber \\
&-  \frac{q  \mbox{ } \left(\mathbf{\hat{z}} \times \mathbf{\hat{t}}\right)\left.\frac{\partial V_{tr}}{\partial z}\right|_{\mathbf{r}=\mathbf{r_{0}}}}{2 g \left|\Psi_{TF}\right|^{2} }  \cdot \left[\ln(\rho) +\mathbf{\hat{\rho}} \otimes \mathbf{\hat{\rho}} \right] \nonumber \\
&+ \mathbf{E}, \label{gsci}
\end{align}
as for large $\rho$, $\frac{\mathbf{r}}{\rho} \approx \mathbf{\hat{\rho}}$ and $\nabla \mathbf{r}$ = $\mathbf{I}$ (where $\mathbf{I}$ is the identity matrix).
 
Matching Eqs.~\eqref{gsci} and \eqref{gsco}, at the vortex core radius $r_{c}$, the equation for $\mathbf{V_{c}}$ becomes 
\begin{align}
\mathbf{V_{c}}(\mathbf{r_{0}}) =& \quad \frac{\hbar q \mbox{ } \kappa  \mathbf{\hat{b}}}{2 M } \cdot \left[\ln( r_{c}) +  \mathbf{\hat{\rho}} \otimes \mathbf{\hat{\rho}}\right]  \nonumber \\
&+\frac{\hbar q \mbox{ } \mathbf{\hat{z}} \times \nabla V_{tr}(\mathbf{r_{0}}) }{2 M g \left|\Psi_{TF}\right|^{2} }  \cdot \left[\ln( r_{c})  +  \mathbf{\hat{\rho}} \otimes \mathbf{\hat{\rho}}\right] \nonumber \\
&+\frac{\hbar q \left(\mathbf{\hat{z}} \times \mathbf{\hat{t}}\right)\left.\frac{\partial V_{tr}}{\partial z}\right|_{\mathbf{r}=\mathbf{r_{0}}}}{2 M g \left|\Psi_{TF}\right|^{2} } \cdot \left[\ln( r_{c})  +  \mathbf{\hat{\rho}} \otimes \mathbf{\hat{\rho}}\right]  \nonumber \\
& +\mathbf{E'}\left[\kappa \mathbf{\hat{b}}, \frac{\mathbf{\hat{z}} \times \nabla V_{tr}(\mathbf{r_{0}})}{g \left|\Psi_{TF}\right|^{2}}, \frac{\mathbf{\hat{z}} \times \mathbf{\hat{t}}}{g \left|\Psi_{TF}\right|^{2}}\left. \frac{\partial V_{tr}}{\partial z}\right|_{\mathbf{r}=\mathbf{r_{0}}}, \mathbf{r_{0}} \right] \nonumber \\
 &- \frac{\hbar q \ln\left(\frac{e^{c } r_{c}}{2 R_{\perp} }\right)}{ M g |\Psi_{TF}|^{2}} [\mathbf{\hat{z}} \times \nabla V_{tr}(\mathbf{r_{0}})],\label{vc}
\end{align}
where $\mathbf{E'} = - \hbar \mathbf{E} / M$.

Combining $\mathbf{V_{c}}$ [Eq.~\eqref{vc}] and $\mathbf{V_{\Omega}}$ [Eq.~\eqref{vo}] the relation for velocity of the vortex line is
\begin{align}
\mathbf{V}(\mathbf{r_{0}}) =& \quad \frac{\hbar q \mbox{ } \kappa  \mathbf{\hat{b}}}{2 M } \cdot \left[\ln( r_{c}) +  \mathbf{\hat{\rho}} \otimes \mathbf{\hat{\rho}}\right]  \nonumber \\
&+\frac{\hbar q \mbox{ } \mathbf{\hat{z}} \times \nabla V_{tr}(\mathbf{r_{0}}) }{2 M g \left|\Psi_{TF}\right|^{2} }  \cdot \left[\ln( r_{c})  +  \mathbf{\hat{\rho}} \otimes \mathbf{\hat{\rho}}\right] \nonumber \\
&+\frac{\hbar q \left(\mathbf{\hat{z}} \times \mathbf{\hat{t}}\right)\left.\frac{\partial V_{tr}}{\partial z}\right|_{\mathbf{r}=\mathbf{r_{0}}}}{2 M g \left|\Psi_{TF}\right|^{2} } \cdot \left[\ln( r_{c})  +  \mathbf{\hat{\rho}} \otimes \mathbf{\hat{\rho}}\right]  \nonumber \\
& +\mathbf{E'}\left[\kappa \mathbf{\hat{b}}, \frac{\mathbf{\hat{z}} \times \nabla V_{tr}(\mathbf{r_{0}})}{g \left|\Psi_{TF}\right|^{2}}, \frac{\mathbf{\hat{z}} \times \mathbf{\hat{t}}}{g \left|\Psi_{TF}\right|^{2}}\left. \frac{\partial V_{tr}}{\partial z}\right|_{\mathbf{r}=\mathbf{r_{0}}}, \mathbf{r_{0}} \right] \nonumber \\
 &- \frac{\hbar q \ln\left(\frac{e^{c } r_{c}}{2 R_{\perp} }\right)}{ M g |\Psi_{TF}|^{2}} [\mathbf{\hat{z}} \times \nabla V_{tr}(\mathbf{r_{0}})] + 2 \frac{\nabla V_{tr} (\mathbf{r_{0}}) \times \mathbf{\Omega} }{\nabla_{\perp}^{2} V_{tr}(\mathbf{r_{0}})},\nonumber \\ \label{Vf}
\end{align}
where $\mathbf{\hat{\rho}}$ is effectively the $\rho$ vector in the condensates coordinates (the difference between $\mathbf{\hat{\rho}}$ of the vortex line and $\mathbf{\hat{\rho}}$ of the condensate is negligible at large distances).

This equation indicates how the local velocity of the vortex line behaves in response to the lines position and curvature, and therefore reveals the local behaviour of a vortex line.

Equation \eqref{Vf} also transforms to give a system of differential equations that describe the complete vortex behaviour.
As the vortex line is a one to one function in $z$, the vortex line's shape and motion are fully parametrized through two functions, $\rho(t,z)$ and $\phi(t,z)$.  This allows any point on the line to be represented by $\mathbf{r_{0}} =  \rho(t,z) \hat{\rho} + \phi(t,z) \hat{\phi} + z \mathbf{\hat{z}}$, the velocity and tangent vectors to be
\begin{eqnarray}
\mbox{Velocity} &= \mathbf{V} & = \frac{\partial \rho}{\partial t} \mathbf{\hat{\rho}} + \rho \frac{\partial \phi}{\partial t} \mathbf{\hat{\phi}},\\
\mbox{tangent vector}& = \mathbf{\hat{t}}& = \frac{\partial \rho}{\partial z} \mathbf{\hat{\rho}} +  \rho \frac{\partial \phi}{\partial z}  \mathbf{\hat{\phi}} + \mathbf{\hat{z}}, \nonumber \\
\end{eqnarray}
and $\kappa \mathbf{\hat{b}}$ ($\approx \mathbf{\hat{z}} \times \mathbf{k}$, where $\mathbf{k} = \partial_{s}^{2} \mathbf{r_{0}}$)  to equal
\begin{eqnarray}
\kappa \mathbf{\hat{b}} =  &-&  \left( \rho \frac{\partial^{2} \phi}{\partial z^{2}} + 2 \frac{\partial \rho}{\partial z}  \frac{\partial \phi}{\partial z} \right) \mathbf{\hat{\rho}} \nonumber \\ 
&+& \left[\frac{\partial^{2} \rho}{\partial z^{2}} - \rho \left(\frac{\partial \phi}{\partial z}\right)^{2} \right] \mathbf{\hat{\phi}}.
\end{eqnarray}
where $\rho$ and $\phi$ are implicitly functions of $t$ and $z$, and $z$ has been approximated to be the arclength of the curve.

As a result, the differential equations describing the full vortex's shape and motion are
\begin{align}
 \frac{\partial \rho}{\partial t} =&- \frac{\hbar q \left[\ln( r_{c})+1\right]}{2 M  }  \left( \rho \frac{\partial^{2} \phi}{\partial z^{2}} + 2 \frac{\partial \rho}{\partial z}  \frac{\partial \phi}{\partial z} \right)\nonumber \\
 &- \frac{\hbar q \left[\ln( r_{c})+1\right]}{2 M g \left|\Psi_{TF}\right|^{2} } \left( \frac{1}{\rho} \frac{\partial V_{tr}}{\partial \phi} + \rho \frac{\partial \phi}{\partial z} \frac{\partial V_{tr}}{\partial z} \right)\nonumber \\
 &+  \frac{2 \Omega }{\rho \nabla_{\perp}^{2} V_{tr}} \frac{\partial V_{tr}}{\partial \phi}  + \frac{\hbar q \ln\left(\frac{e^{c } r_{c}}{2 R_{\perp} }\right)}{ \rho M g |\Psi_{TF}|^{2}} \frac{\partial V_{tr}}{\partial \phi} \nonumber\\
 &+\mathbf{\hat{\rho}}\cdot\mathbf{E'}, \label{rho}\\ 
 \nonumber \\
\rho \frac{\partial \phi}{\partial t} =&\quad \frac{\hbar q \ln( r_{c})}{2 M  }  \left[\frac{\partial^{2} \rho}{\partial z^{2}} - \rho \left(\frac{\partial \phi}{\partial z}\right)^{2} \right] \nonumber \\
&+ \frac{\hbar q \ln( r_{c})}{2 M g \left|\Psi_{TF}\right|^{2} } \left( \frac{\partial V_{tr}}{\partial \rho} + \frac{\partial \rho}{\partial z} \frac{\partial V_{tr}}{\partial z} \right)\nonumber \\
 & -  \frac{2 \Omega }{\nabla_{\perp}^{2} V_{tr}} \frac{\partial V_{tr}}{\partial \rho}
 - \frac{\hbar q \ln\left(\frac{e^{c } r_{c}}{2 R_{\perp} }\right)}{ M g |\Psi_{TF}|^{2}} \frac{\partial V_{tr}}{\partial \rho}  \nonumber \\
&+ \hat{\phi}\cdot\mathbf{E'}. \label{phi}
\end{align}

The solutions of Eqs.~\eqref{rho} and \eqref{phi} describes a vortex's shape and motion, and shows the steady state structures and waves the vortex supports. However, before these structures are determined the unknown constant $\mathbf{E'}$ needs to be found.

\section{Determining $\mathbf{E'}$} \label{sec:DE} 

Before investigating Eqs.~\eqref{rho} and \eqref{phi} $\mathbf{E'}$ needs to be determined. This can be done through comparing specific results from these equations to a previously determined physical scenarios.

Assuming that the trapping and curving dependence in $\mathbf{E'}$ can be treated separately, two different scenarios need to be considered: one indicting a straight vortex's reaction to the trap confinement, and one describing behaviour of wave perturbations on the line in the absence of confinement.

In a pancake shaped condensate it has been shown that the vortex lines are straight, while for more cigar shaped traps the line tends to bend at the edges \cite{PhysRevA.64.043611}. Such straight lines are easy to work with mathematically and so the energetics and behaviour of these lines have been investigated \cite{PhysRevLett.84.5919, PhysRevA.63.043608}. These investigations revealed that a off centred straight line vortex precesses around a cylindrical condensate in the direction of the vortex's rotation with a precession frequency
\begin{equation} \label{pres}
\frac{\partial \phi}{\partial t} = \frac{3 \hbar \ln\left(\frac{R_{\perp}}{r_{c}}\right)}{2 M R_{\perp}^{2} \left(1 - \frac{\rho^{2}}{R_{\perp}^{2}}\right)} - \Omega.
\end{equation}

Applying similar conditions to Eqs.~\eqref{rho} and \eqref{phi} (discarding $z$ derivatives and setting $V_{tr} = M \omega_{\perp}^{2} \rho^{2}/2 + M \omega_{z}^{2} z^{2}/2$) and setting $g |\Psi_{TF}|^{2}$ $\approx$ $\frac{1}{2} M \omega_{\perp}^{2} R_{\perp}^{2} \left(1 - \frac{\rho^{2}}{R_{\perp}^{2}}\right)$ the describing equations become
\begin{align}
 \frac{\partial \rho}{\partial t} =& \quad \mathbf{\hat{\rho}}\cdot\mathbf{E'}\left[0,\frac{2 \rho \hat{\phi}}{R_{\perp}^{2} \left( 1 - \frac{\rho^{2}}{R_{\perp}^{2}}\right)},0,\mathbf{r}\right], \\
\rho \frac{\partial \phi}{\partial t} =& - \frac{\hbar \ln\left( \frac{e^{2c } r_{c}}{4 R_{\perp}^{2} }\right) \rho}{ M R_{\perp}^{2} \left( 1 - \frac{\rho^{2}}{R_{\perp}^{2}}\right)}  -  \Omega \rho \nonumber \\
& + \hat{\phi}\cdot\mathbf{E'} \left[0,\frac{2 \rho \hat{\phi}}{R_{\perp}^{2} \left( 1 - \frac{\rho^{2}}{R_{\perp}^{2}}\right)},0,\mathbf{r}\right].
\end{align}

When matched to Eq. \eqref{pres} these equations become
\begin{gather}
\mathbf{\hat{\rho}}\cdot\mathbf{E'}\left[0,\frac{2 \rho \hat{\phi}}{R_{\perp}^{2} \left( 1 - \frac{\rho^{2}}{R_{\perp}^{2}}\right)},0,\mathbf{r}\right] = 0, \\ \nonumber \\
 \hat{\phi}\cdot\mathbf{E'} \left[0,\frac{2 \rho \hat{\phi}}{R_{\perp}^{2} \left( 1 - \frac{\rho^{2}}{R_{\perp}^{2}}\right)},0,\mathbf{r}\right] =  \frac{\hbar \ln\left( \frac{e^{2c }}{4 \sqrt{R_{\perp} r_{c}} }\right) \rho}{ M R_{\perp}^{2} \left( 1 - \frac{\rho^{2}}{R_{\perp}^{2}}\right)}.
\end{gather}

This suggests that the trap dependence in the constant term has the form
\begin{gather}
\mathbf{E'}\left(\kappa \mathbf{\hat{b}}, \frac{\mathbf{\hat{z}} \times \nabla V_{tr}}{g \left|\Psi_{TF}\right|^{2}}, \frac{\mathbf{\hat{z}} \times \mathbf{\hat{t}}}{g \left|\Psi_{TF}\right|^{2}} \frac{\partial V_{tr}}{\partial z}, \mathbf{r} \right) \nonumber \\
= \frac{\hbar q  \ln\left( \frac{e^{2c }}{4 \sqrt{R_{\perp} r_{c}} }\right)\left(\mathbf{\hat{z}} \times \nabla V_{tr}\right)}{2 M g \left|\Psi_{TF}\right|^{2} }  + \mathbf{F}\left(\kappa \mathbf{\hat{b}}, \mathbf{r} \right), \nonumber \\
\end{gather}
where $\mathbf{F}$ is an unknown constant that describes the curvature behaviour.

Note that this condition, though intending to contain all trap dependence, does not contain information on the lines reaction to the $z$ trapping of the condensate. This indicates that another condition relating specifically to this term is necessary (visible through the $\ln$ term). Unfortunately a suitable comparison was not found and therefore this term could not be calibrated.

To determine the curvature related behaviour of the constant, the behaviour of a perturbed vortex (in a untrapped condensate) needs to be known. It has been shown, theoretically, that vortices in BECs support helical wave perturbations \cite{Sov.Phys.JETP.13.451} and therefore are an ideal example. These waves rotate in the opposite direction to the rotation of the vortex, and in a uniform condensate has a dispersion relation (for large wavelengths) of
\begin{equation} \label{pit}
\omega = \frac{\hbar k^{2}}{2 M} \ln\left(\frac{1}{|k| r_{c}}\right),
\end{equation}
where $\omega$ is the wave frequency, $k$ is the wave number and $r_{c}$ is the vortex core radius.
Hence by comparing this dispersion relation to the equivalent derived from Eqs.~\eqref{rho} and \eqref{phi} the form of $\mathbf{F}$ can be determined.

A helical wave generally has the form $ x = a \sin(k z -\omega t +\phi_{0})$ and $ y = a \cos(k z -\omega t +\phi_{0})$, which when transformed into cylindrical coordinates is $ \rho = a $ and $\phi = k z -\omega t +\phi_{0}$ ($a$ being the wave amplitude and $\phi_{0}$ being an arbitrary phase constant). Substituting this parametrization into Eqs.~\eqref{rho} and \eqref{phi}, they become

\begin{equation}
 0 =- \frac{\hbar [\ln( r_{c})+1]}{2 M g \left|\Psi_{TF}\right|^{2} } \left( 0 \right) +
   \mathbf{\hat{\rho}}\cdot\mathbf{F}\left(-a k^{2} \hat{\phi},\mathbf{r}\right)
\end{equation}
and
\begin{equation}
- a \omega = - \frac{\hbar a k^{2} \ln( r_{c})}{2 M } +
 \hat{\phi}\cdot\mathbf{F}\left(-a k^{2} \hat{\phi},\mathbf{r}\right) .
\end{equation}

Hence, the conditions $\mathbf{F}$ satisfies are
\begin{gather}
\mathbf{\hat{\rho}}\cdot\mathbf{F}\left(-a k^{2} \hat{\phi},\mathbf{r}\right) =0, \\
  \hat{\phi}\cdot\mathbf{F}\left(-a k^{2} \hat{\phi},\mathbf{r}\right)=a \frac{\hbar k^{2}}{2 M} \ln(|k| r_{c}^{2}).
\end{gather}

Since all the wavenumber dependence in $\mathbf{F}$ must come from the $\kappa \mathbf{\hat{b}}$ term a suitable form becomes
\begin{equation}
\mathbf{F} =  \frac{\hbar q \ln \left( \frac{1}{r_{c}^{2}} \sqrt{\frac{- \rho}{\kappa \mathbf{\hat{b}}\cdot\hat{\phi}}} \right)}{2 M}   \kappa \mathbf{\hat{b}}.
\end{equation}

Substituting the full form of $\mathbf{E'}$ into Eq. \eqref{Vf} the velocity condition becomes
\begin{align}
\mathbf{V}(\mathbf{r_{0}})  = &  \quad \frac{\hbar q  \left(\mathbf{\hat{z}} \times \mathbf{\hat{t}}\right) \left.\frac{\partial V_{tr}}{\partial z}\right|_{\mathbf{r}=\mathbf{r_{0}}}}{2 M g \left|\Psi_{TF}\right|^{2} }  \cdot \left[\ln( r_{c}) + \mathbf{\hat{\rho}} \otimes \mathbf{\hat{\rho}} \right] \nonumber \\ 
&+\frac{\hbar q \mbox{ } \kappa \mathbf{\hat{b}} }{2 M}   \cdot\left[\ln \left( \frac{1}{r_{c}} \sqrt{\frac{- \rho_{0}}{\kappa \mathbf{\hat{b}}\cdot\hat{\phi}} }\right) + \mathbf{\hat{\rho}} \otimes \mathbf{\hat{\rho}}\right] \nonumber \\ 
&+ \frac{\hbar q  \left(\mathbf{\hat{z}} \times \nabla V_{tr}(\mathbf{r_{0}})\right) }{2 M g \left|\Psi_{TF}\right|^{2} } \cdot \left[ \frac{3}{2}\ln\left( \frac{R_{\perp}}{ r_{c}} \right) + \mathbf{\hat{\rho}} \otimes \mathbf{\hat{\rho}} \right] \nonumber \\
&+ 2 \frac{\nabla V_{tr} (\mathbf{r_{0}}) \times \mathbf{\Omega} }{\nabla_{\perp}^{2} V_{tr}(\mathbf{r_{0}})}.  \label{Vf2}
\end{align}

This is an almost complete relation for the velocity of the vortex line, missing the calibration of the $\partial_{z} V_{tr}$ term [as evidenced by the units in the multiplying logarithm $\ln(r_{c})$]. As a suitable analytic form for this comparison could not be found, and it does not affect any of the latter calculations performed, this issue will be left for future work.

 When transformed into differential form (as per Section \ref{sec:match}), this equation gives
\begin{align}
\frac{\partial \rho}{\partial t}  = & - \frac{\hbar q \rho \left[\ln( r_{c})+1\right] }{2  M g \left|\Psi_{TF}\right|^{2} }  \frac{\partial \phi}{\partial z} \frac{\partial V_{tr}}{\partial z}  \nonumber \\ 
&-\frac{\hbar q \left[\ln \left( \frac{1}{r_{c}} \sqrt{\frac{ \rho}{\rho \left(\partial_{z} \phi\right)^{2} - \partial_{z}^{2} \rho} }\right)+1\right] }{2 M} \left( \rho \frac{\partial^{2} \phi}{\partial z^{2}} + 2 \frac{\partial \rho}{\partial z}  \frac{\partial \phi}{\partial z} \right) \nonumber \\ 
&- \frac{3\hbar q  \left[\ln\left( \frac{R_{\perp}}{ r_{c}} \right)+\frac{2}{3}\right] \mbox{ }  }{4 \rho M g \left|\Psi_{TF}\right|^{2} } \frac{\partial V_{tr}}{\partial \phi} \nonumber \\ 
&+ 2 \frac{\Omega }{\rho \nabla_{\perp}^{2} V_{tr}} \frac{\partial V_{tr}}{\partial \phi},  \label{rho2}
\end{align}
and
\begin{align}
\rho \frac{\partial\phi}{\partial t}  = & \quad \frac{\hbar q \ln( r_{c})}{2 M g \left|\Psi_{TF}\right|^{2} }  \frac{\partial \rho}{\partial z} \frac{\partial V_{tr}}{\partial z} + \frac{3\hbar q  \ln\left( \frac{R_{\perp}}{ r_{c}} \right)  }{4 M g \left|\Psi_{TF}\right|^{2} } \frac{\partial V_{tr}}{\partial \rho} \nonumber \\ 
&+\frac{\hbar q \ln \left( \frac{1}{r_{c}} \sqrt{\frac{ \rho}{\rho \left(\partial_{z} \phi\right)^{2} - \partial_{z}^{2} \rho} }\right)}{2 M}  \left[\frac{\partial^{2} \rho}{\partial z^{2}} - \rho \left(\frac{\partial \phi}{\partial z}\right)^{2} \right] \nonumber \\ 
&- 2 \frac{\Omega}{\nabla_{\perp}^{2} V_{tr}} \frac{\partial V_{tr}}{\partial \rho}. \label{phi2}
\end{align}

From these non-linear equations the complete behaviour of the vortex line can be determined, however, this paper will only look at some general and simple examples of these equations.

\section{wave investigations} \label{sec:waves}
\subsection{Qualitative comments}  
These equations give deterministic properties of a slightly perturbed vortex in a BEC. However before looking at a special case of this, it is worth looking at qualitative behaviour from Eq. \eqref{Vf2}.

The vortex contains angular momentum in $\mathbf{\hat{z}}$. Therefore, as suggested by A. Fetter \cite{RevModPhys.81.647}, the vortex can be considered as a gyroscope spinning in $\mathbf{\hat{z}}$, responding to any force by moving perpendicular to it. In Eq. \eqref{Vf2} there are 3 such forces: a trap gradient force ($-\nabla V_{tr}(\mathbf{r_{0}})$), a curvature force ($\kappa \mathbf{\hat{n}}$, $\mathbf{\hat{b}} \approx \mathbf{\hat{z}} \times \mathbf{\hat{n}}$), and a z-trap deviation force ($\partial_{z} V_{tr}(\mathbf{r_{0}}) \mathbf{\hat{t}}$). Each force contributing to the relations final form.

Similarly, the rotation of the condensate $\mathbf{\Omega}$, also creates gyroscopic motion. This rotation causes any point in the trap to move perpendicular to the trap gradient force, and therefore also pushes the vortex line accordingly. 

Equation \eqref{Vf2} also contains a collection of terms that `amplifies' the velocity of the line in the $\rho$ direction. These terms are additional restoring forces, which attempt to return the system to the the centred straight line vortex state.

Hence the qualitative behaviour depicted by Eq.\eqref{Vf2} gives understandable insight into how the vortex line moves; therefore, indicating that it is practical to use it to quantitatively determine the structure and motion of the vortex line in a condensate.

\subsection{More realistic helical wave behaviour} \label{sec:Wshw} 

As previously mentioned, helical waves can exist on vortex lines in a BEC. Hence it is relevant to consider the behaviour of such waves in the presence of a trapping potential. Assuming a cylindrical trap ($V_{tr} = M \omega_{\perp}^{2} \rho^{2} /2$, where $\omega_{z}=0$) and that the helix wave takes the same formed used in Section \ref{sec:DE} ($\rho=a$ and $\phi = k z - \omega t + \phi_{0}$), Eq.~\eqref{rho2} is solved and Eq.~\eqref{phi2} becomes
\begin{eqnarray}
\omega  = - \frac{3\hbar q  \ln\left( \frac{R_{\perp}}{ r_{c}} \right)  }{2 M R_{\perp}^{2} \left(1 - \frac{\rho^{2}}{R_{\perp}^{2}}\right) }+\frac{\hbar q  k^{2} \ln \left( \frac{1}{r_{c} |k|} \right)}{2 M}   + \Omega . \label{dis}
\end{eqnarray}
This equation shows that the precession helical wave in a cylindrical trap is a combination of the precession of a straight line vortex and the motion of a helical wave in an uniform condensate.

\begin{figure}[ht]
\centering
\includegraphics[width=0.4\textwidth]{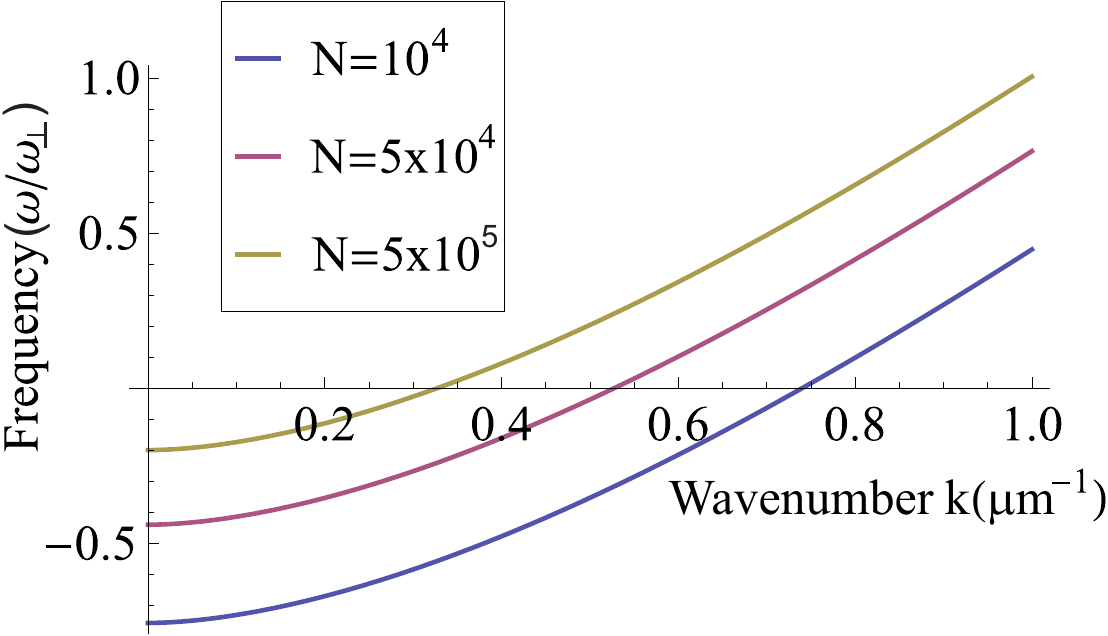}
\caption{Plots of the dispersion relation, Eq.~\eqref{dis}, for different particle number ($N$) for a condensate of $^{87}Rb$ atoms, with trapping frequencies $\omega_{\perp} = 2 \pi \times 11.8 \mbox{ Hz} $, and $\omega_{z} =  2 \pi \times 98.5 \mbox{ Hz}$, and vortex core radius $r_{c} = 0.13 \mu m$.}
\label{fig:tap}
\end{figure}

 In a numerical simulation by T. Simula \textit{et al.} \cite{PhysRevLett.101.020402, PhysRevA.78.053604}, it was shown that the dispersion relation of large wavelength helical waves in a cigar shaped trap could be represented as
\begin{equation}
\omega  = \omega_{0} +\frac{\hbar q  k^{2} \ln \left(\frac{1}{r_{c} |k|} \right)}{2 M}   + \Omega,
\end{equation}
where $\omega_{0}$ is a constant precession frequency.
The similarity of these two equations is striking, and upon plotting (Fig.~\ref{fig:tap}), the similarities are still visible (in the change to $\omega$ to the changing particle number), however the specific values of the constant vary by about $\pm 0.1 (\omega/\omega_{\perp})$ to the numerical value.

This deviation is not surprising. Equation \eqref{dis}'s constant precession is that of a purely straight line vortex. This suggests that Eq.~\eqref{dis} is representative of pancake shaped traps where such vortices occur. T. Simula's work however is in a cigar shaped trap and therefore the vortex line, without wave perturbations, tends to bend, potentially changing its precession. Therefore to accurately depict this behaviour the trapping of the condensate in $z$, and how it distorts the vortex, needs to be considered.

\section{Previous similar work} \label{sec:Psw}
A similar derivation has been carried out by A. Svidzinsky and A. Fetter \cite{Phys.Rev.A.62.063617} with the aim of determining the normal waves that a vortex line can support. This work performed a similar matched asymptotic expansion to determine a relation for the vortex lines velocity.

During this derivation, an extra term appeared in the outer equations that was related to the transformation into the local coordinates of the vortex line, the specifics of which were not explained in the paper. This gave an outer solution of
\begin{equation} \label{Fo}
\Phi = - e^{\frac{\kappa x}{2}} K_{0} \left( \rho \sqrt{\frac{ \nabla_{\perp}^{2} V_{tr} }{2 g |\Psi_{TF}|^{2}}+ \frac{\kappa^{2}}{4}} \right),
\end{equation}
which was approximated to be
\begin{equation}
\Phi = \ln\left( \frac{e^{c}}{\sqrt{2}} \rho \sqrt{\frac{ 1 }{R_{\perp}^{2}}+ \frac{\kappa^{2}}{4}} \right),
\end{equation}
for small $\rho$. This form is almost identical to that derived in this paper, having effectively ignored the additional exponential in their outer solution. 
 
Then to compare the outer and inner solutions a pseudo-$\Phi$ was created for the inner phase solution. This pseudo-$\Phi$ was linked with the inner solution phase $S$ through the same definition of $\Phi$ used in Section \ref{sec:outer}, but ignored some terms, of comparable order to those kept,
\begin{equation}
\Phi \approx \left[ 1 + \frac{x}{2} \left( \frac{|\nabla_{\perp} V_{tr}|}{ g |\Psi_{TF}|^{2}} + \kappa \right) \right] \ln( E \rho) + \frac{M V_{0}}{\hbar q} x.
\end{equation}
It is worth noting that the largest term of this inner $\Phi$ is $x \ln(\rho)$, while the largest term of the outer $\Phi$ is $\ln(\rho)$. This indicates an inconsistency in the expansion of the outer solution (the e$^\frac{\kappa x}{2}$ in Eq.~\eqref{Fo} allowing the outer to also expand to $x \ln(\rho)$). 

Through matching these $\Phi$s the velocity condition found was
\begin{eqnarray}
\mathbf{V(r_{0})} =& -&\frac{q \hbar \mathbf{\hat{t}} \times \nabla V_{tr}}{2 M g |\Psi_{TF}|^{2}} \ln\left( |q| r_{c} \sqrt{\frac{ 1 }{R_{\perp}^{2}}+ \frac{\kappa^{2}}{4}} \right)\nonumber \\
&-& \frac{q \hbar  \kappa \mathbf{\hat{b}} }{2 M}  \ln\left( |q| r_{c} \sqrt{\frac{ 1 }{R_{\perp}^{2}}+ \frac{\kappa^{2}}{4}} \right)\nonumber \\
 &+& 2 \frac{   \nabla V_{tr} \times \mathbf{\Omega}}{ \nabla^{2} V_{tr}}.
\end{eqnarray}
This condition gives similar wave results within certain limits. Specifically both equations replicate the long wavelength helical wave dispersion relation but generally give different results and dispersion relations. 

Furthermore, in their derivation the effect of $\mathbf{\hat{t}}$ deviating from $\mathbf{\hat{z}}$ was not considered, but later implemented through the $\mathbf{\hat{t}} \times \nabla V_{tr}$ term. This procedure gives the same $\partial_{z} V_{tr}$ structure seen in Eq. \eqref{Vf2} however also adds an additional artefact in the $\mathbf{\hat{z}}$ direction of the velocity equation.

\section{Conclusions and Future Directions} \label{sec:VLC}
Through this derivation, a robust relation for the motion and shape of a vortex line in a BEC was determined [Eq.~\eqref{Vf2}]. This relation was found by determining the behaviour of the condensate near and far from the vortex core separately, and ensuring that these two solutions matched in the overlapping region. 

This relation was shown to describe well the local vortex's behaviour, and could be transformed into a set of differential equation that describes the vortex's overall structure and motion [Eqs.~\eqref{rho2} and \eqref{phi2}]. Using a simple calculation, these equations reflected wave behaviour seen numerically.

This paper only briefly explored the behaviour of Eqs.~\eqref{rho2} and \eqref{phi2}, and so further investigation needs to be performed, looking at different vortex line structures and trapping geometries. However before this can be done an analytic description of how the trapping in $z$ affects the line's behaviour is needed to calibrate the final term in Eq.~\eqref{Vf2}.

\section*{Acknowledgements}
Thanks to Alexander Fetter and Tapio Simula for useful discussions and general advice.

\bibliography{Bibliography}

\begin{thebibliography}{30}
\expandafter\ifx\csname natexlab\endcsname\relax\def\natexlab#1{#1}\fi
\expandafter\ifx\csname bibnamefont\endcsname\relax
  \def\bibnamefont#1{#1}\fi
\expandafter\ifx\csname bibfnamefont\endcsname\relax
  \def\bibfnamefont#1{#1}\fi
\expandafter\ifx\csname citenamefont\endcsname\relax
  \def\citenamefont#1{#1}\fi
\expandafter\ifx\csname url\endcsname\relax
  \def\url#1{\texttt{#1}}\fi
\expandafter\ifx\csname urlprefix\endcsname\relax\def\urlprefix{URL }\fi
\providecommand{\bibinfo}[2]{#2}
\providecommand{\eprint}[2][]{\url{#2}}

\bibitem[{\citenamefont{Matthews et~al.}(1999)\citenamefont{Matthews, Anderson,
  Haljan, Hall, Wieman, and Cornell}}]{PhysRevLett.83.2498}
\bibinfo{author}{\bibfnamefont{M.~R.} \bibnamefont{Matthews}},
  \bibinfo{author}{\bibfnamefont{B.~P.} \bibnamefont{Anderson}},
  \bibinfo{author}{\bibfnamefont{P.~C.} \bibnamefont{Haljan}},
  \bibinfo{author}{\bibfnamefont{D.~S.} \bibnamefont{Hall}},
  \bibinfo{author}{\bibfnamefont{C.~E.} \bibnamefont{Wieman}},
  \bibnamefont{and} \bibinfo{author}{\bibfnamefont{E.~A.}
  \bibnamefont{Cornell}}, \bibinfo{journal}{Phys. Rev. Lett.}
  \textbf{\bibinfo{volume}{83}}, \bibinfo{pages}{2498} (\bibinfo{year}{1999}).

\bibitem[{\citenamefont{Anderson et~al.}(2000)\citenamefont{Anderson, Haljan,
  Wieman, and Cornell}}]{PhysRevLett.85.2857}
\bibinfo{author}{\bibfnamefont{B.~P.} \bibnamefont{Anderson}},
  \bibinfo{author}{\bibfnamefont{P.~C.} \bibnamefont{Haljan}},
  \bibinfo{author}{\bibfnamefont{C.~E.} \bibnamefont{Wieman}},
  \bibnamefont{and} \bibinfo{author}{\bibfnamefont{E.~A.}
  \bibnamefont{Cornell}}, \bibinfo{journal}{Phys. Rev. Lett.}
  \textbf{\bibinfo{volume}{85}}, \bibinfo{pages}{2857} (\bibinfo{year}{2000}).

\bibitem[{\citenamefont{Madison et~al.}(2000)\citenamefont{Madison, Chevy,
  Wohlleben, and Dalibard}}]{PhysRevLett.84.806}
\bibinfo{author}{\bibfnamefont{K.~W.} \bibnamefont{Madison}},
  \bibinfo{author}{\bibfnamefont{F.}~\bibnamefont{Chevy}},
  \bibinfo{author}{\bibfnamefont{W.}~\bibnamefont{Wohlleben}},
  \bibnamefont{and} \bibinfo{author}{\bibfnamefont{J.}~\bibnamefont{Dalibard}},
  \bibinfo{journal}{Phys. Rev. Lett.} \textbf{\bibinfo{volume}{84}},
  \bibinfo{pages}{806} (\bibinfo{year}{2000}).

\bibitem[{\citenamefont{Chevy et~al.}(2000)\citenamefont{Chevy, Madison, and
  Dalibard}}]{PhysRevLett.85.2223}
\bibinfo{author}{\bibfnamefont{F.}~\bibnamefont{Chevy}},
  \bibinfo{author}{\bibfnamefont{K.~W.} \bibnamefont{Madison}},
  \bibnamefont{and} \bibinfo{author}{\bibfnamefont{J.}~\bibnamefont{Dalibard}},
  \bibinfo{journal}{Phys. Rev. Lett.} \textbf{\bibinfo{volume}{85}},
  \bibinfo{pages}{2223} (\bibinfo{year}{2000}).

\bibitem[{\citenamefont{Bretin et~al.}(2004)\citenamefont{Bretin, Stock,
  Seurin, and Dalibard}}]{PhysRevLett.92.050403}
\bibinfo{author}{\bibfnamefont{V.}~\bibnamefont{Bretin}},
  \bibinfo{author}{\bibfnamefont{S.}~\bibnamefont{Stock}},
  \bibinfo{author}{\bibfnamefont{Y.}~\bibnamefont{Seurin}}, \bibnamefont{and}
  \bibinfo{author}{\bibfnamefont{J.}~\bibnamefont{Dalibard}},
  \bibinfo{journal}{Phys. Rev. Lett.} \textbf{\bibinfo{volume}{92}},
  \bibinfo{pages}{050403} (\bibinfo{year}{2004}).

\bibitem[{550(2010)}]{5502338220101201}
\bibinfo{journal}{Journal of Low Temperature Physics}
  \textbf{\bibinfo{volume}{161}}, \bibinfo{pages}{574} (\bibinfo{year}{2010}),
  ISSN \bibinfo{issn}{00222291}.

\bibitem[{\citenamefont{Fetter}(2009)}]{RevModPhys.81.647}
\bibinfo{author}{\bibfnamefont{A.~L.} \bibnamefont{Fetter}},
  \bibinfo{journal}{Rev. Mod. Phys.} \textbf{\bibinfo{volume}{81}},
  \bibinfo{pages}{647} (\bibinfo{year}{2009}).

\bibitem[{\citenamefont{Feynman}(1955)}]{Prog.LowTemp.Phys.1.17.}
\bibinfo{author}{\bibfnamefont{R.~P.} \bibnamefont{Feynman}},
  \bibinfo{journal}{Prog. Low Temp. Phys.} \textbf{\bibinfo{volume}{1}},
  \bibinfo{pages}{17} (\bibinfo{year}{1955}).

\bibitem[{\citenamefont{Pethick and Smith}(2002)}]{PethickSmith}
\bibinfo{author}{\bibfnamefont{C.}~\bibnamefont{Pethick}} \bibnamefont{and}
  \bibinfo{author}{\bibfnamefont{H.}~\bibnamefont{Smith}},
  \emph{\bibinfo{title}{Bose-Einstein Condensation in Dilute Gases}}
  (\bibinfo{publisher}{Cambridge}, \bibinfo{year}{2002}).

\bibitem[{\citenamefont{Pitaevskii}(1961)}]{Sov.Phys.JETP.13.451}
\bibinfo{author}{\bibfnamefont{L.~P.} \bibnamefont{Pitaevskii}},
  \bibinfo{journal}{Sov. Phys. JETP} \textbf{\bibinfo{volume}{13}},
  \bibinfo{pages}{451} (\bibinfo{year}{1961}).

\bibitem[{\citenamefont{Gross}(1961)}]{NuovoCimento.20.454}
\bibinfo{author}{\bibfnamefont{E.~P.} \bibnamefont{Gross}},
  \bibinfo{journal}{Nuovo Cimento} \textbf{\bibinfo{volume}{20}},
  \bibinfo{pages}{454} (\bibinfo{year}{1961}).

\bibitem[{\citenamefont{Svidzinsky and Fetter}(2000)}]{PhysRevLett.84.5919}
\bibinfo{author}{\bibfnamefont{A.~A.} \bibnamefont{Svidzinsky}}
  \bibnamefont{and} \bibinfo{author}{\bibfnamefont{A.~L.}
  \bibnamefont{Fetter}}, \bibinfo{journal}{Phys. Rev. Lett.}
  \textbf{\bibinfo{volume}{84}}, \bibinfo{pages}{5919} (\bibinfo{year}{2000}).

\bibitem[{\citenamefont{Fedichev and Shlyapnikov}(1999)}]{PhysRevA.60.R1779}
\bibinfo{author}{\bibfnamefont{P.~O.} \bibnamefont{Fedichev}} \bibnamefont{and}
  \bibinfo{author}{\bibfnamefont{G.~V.} \bibnamefont{Shlyapnikov}},
  \bibinfo{journal}{Phys. Rev. A} \textbf{\bibinfo{volume}{60}},
  \bibinfo{pages}{R1779} (\bibinfo{year}{1999}).

\bibitem[{\citenamefont{McGee and Holland}(2001)}]{PhysRevA.63.043608}
\bibinfo{author}{\bibfnamefont{S.~A.} \bibnamefont{McGee}} \bibnamefont{and}
  \bibinfo{author}{\bibfnamefont{M.~J.} \bibnamefont{Holland}},
  \bibinfo{journal}{Phys. Rev. A} \textbf{\bibinfo{volume}{63}},
  \bibinfo{pages}{043608} (\bibinfo{year}{2001}).

\bibitem[{\citenamefont{Aftalion and Riviere}(2001)}]{PhysRevA.64.043611}
\bibinfo{author}{\bibfnamefont{A.}~\bibnamefont{Aftalion}} \bibnamefont{and}
  \bibinfo{author}{\bibfnamefont{T.}~\bibnamefont{Riviere}},
  \bibinfo{journal}{Phys. Rev. A} \textbf{\bibinfo{volume}{64}},
  \bibinfo{pages}{043611} (\bibinfo{year}{2001}).

\bibitem[{\citenamefont{Garc{\'i}a-Ripoll and
  P{\'e}rez-Garc{\'i}a}(2001)}]{PhysRevA.64.053611}
\bibinfo{author}{\bibfnamefont{J.~J.} \bibnamefont{Garc{\'i}a-Ripoll}}
  \bibnamefont{and} \bibinfo{author}{\bibfnamefont{V.~M.}
  \bibnamefont{P{\'e}rez-Garc{\'i}a}}, \bibinfo{journal}{Phys. Rev. A}
  \textbf{\bibinfo{volume}{64}}, \bibinfo{pages}{053611}
  (\bibinfo{year}{2001}).

\bibitem[{\citenamefont{Rosenbusch et~al.}(2002)\citenamefont{Rosenbusch,
  Bretin, and Dalibard}}]{PhysRevLett.89.200403}
\bibinfo{author}{\bibfnamefont{P.}~\bibnamefont{Rosenbusch}},
  \bibinfo{author}{\bibfnamefont{V.}~\bibnamefont{Bretin}}, \bibnamefont{and}
  \bibinfo{author}{\bibfnamefont{J.}~\bibnamefont{Dalibard}},
  \bibinfo{journal}{Phys. Rev. Lett.} \textbf{\bibinfo{volume}{89}},
  \bibinfo{pages}{200403} (\bibinfo{year}{2002}).

\bibitem[{\citenamefont{Modugno et~al.}(2003)\citenamefont{Modugno,
  Pricoupenko, and Castin}}]{Eur.Phys.J.D.22}
\bibinfo{author}{\bibfnamefont{M.}~\bibnamefont{Modugno}},
  \bibinfo{author}{\bibfnamefont{L.}~\bibnamefont{Pricoupenko}},
  \bibnamefont{and} \bibinfo{author}{\bibfnamefont{Y.}~\bibnamefont{Castin}},
  \bibinfo{journal}{Eur. Phys. J. D} \textbf{\bibinfo{volume}{22}},
  \bibinfo{pages}{235} (\bibinfo{year}{2003}), ISSN \bibinfo{issn}{14346060}.

\bibitem[{\citenamefont{Zambelli and Stringari}(1998)}]{PhysRevLett.81.1754}
\bibinfo{author}{\bibfnamefont{F.}~\bibnamefont{Zambelli}} \bibnamefont{and}
  \bibinfo{author}{\bibfnamefont{S.}~\bibnamefont{Stringari}},
  \bibinfo{journal}{Phys. Rev. Lett.} \textbf{\bibinfo{volume}{81}},
  \bibinfo{pages}{1754} (\bibinfo{year}{1998}).

\bibitem[{\citenamefont{Haljan et~al.}(2001)\citenamefont{Haljan, Anderson,
  Coddington, and Cornell}}]{PhysRevLett.86.2922}
\bibinfo{author}{\bibfnamefont{P.~C.} \bibnamefont{Haljan}},
  \bibinfo{author}{\bibfnamefont{B.~P.} \bibnamefont{Anderson}},
  \bibinfo{author}{\bibfnamefont{I.}~\bibnamefont{Coddington}},
  \bibnamefont{and} \bibinfo{author}{\bibfnamefont{E.~A.}
  \bibnamefont{Cornell}}, \bibinfo{journal}{Phys. Rev. Lett.}
  \textbf{\bibinfo{volume}{86}}, \bibinfo{pages}{2922} (\bibinfo{year}{2001}).

\bibitem[{\citenamefont{{W. Thomson (Lord Kelvin)}}(1880)}]{Philos.mag.10.155}
\bibinfo{author}{\bibnamefont{{W. Thomson (Lord Kelvin)}}},
  \bibinfo{journal}{Philos. mag} \textbf{\bibinfo{volume}{10}},
  \bibinfo{pages}{155} (\bibinfo{year}{1880}).

\bibitem[{\citenamefont{Bretin et~al.}(2003)\citenamefont{Bretin, Rosenbusch,
  Chevy, Shlyapnikov, and Dalibard}}]{PhysRevLett.90.100403}
\bibinfo{author}{\bibfnamefont{V.}~\bibnamefont{Bretin}},
  \bibinfo{author}{\bibfnamefont{P.}~\bibnamefont{Rosenbusch}},
  \bibinfo{author}{\bibfnamefont{F.}~\bibnamefont{Chevy}},
  \bibinfo{author}{\bibfnamefont{G.~V.} \bibnamefont{Shlyapnikov}},
  \bibnamefont{and} \bibinfo{author}{\bibfnamefont{J.}~\bibnamefont{Dalibard}},
  \bibinfo{journal}{Phys. Rev. Lett.} \textbf{\bibinfo{volume}{90}},
  \bibinfo{pages}{100403} (\bibinfo{year}{2003}).

\bibitem[{\citenamefont{Rubinstein and Pismen}(1994)}]{PhysicaD78.1994.1}
\bibinfo{author}{\bibfnamefont{B.~Y.} \bibnamefont{Rubinstein}}
  \bibnamefont{and} \bibinfo{author}{\bibfnamefont{L.~M.}
  \bibnamefont{Pismen}}, \bibinfo{journal}{Physica D}
  \textbf{\bibinfo{volume}{78}}, \bibinfo{pages}{1} (\bibinfo{year}{1994}).

\bibitem[{\citenamefont{Pismen and Rubinstein}(1991)}]{PhysicaD.47.1991.353}
\bibinfo{author}{\bibfnamefont{L.~M.} \bibnamefont{Pismen}} \bibnamefont{and}
  \bibinfo{author}{\bibfnamefont{J.}~\bibnamefont{Rubinstein}},
  \bibinfo{journal}{Physica D} \textbf{\bibinfo{volume}{47}},
  \bibinfo{pages}{353} (\bibinfo{year}{1991}).

\bibitem[{\citenamefont{{Anatoly A. Svidzinsky} and {Alexander L.
  Fetter}}(2000)}]{Phys.Rev.A.62.063617}
\bibinfo{author}{\bibnamefont{{Anatoly A. Svidzinsky}}} \bibnamefont{and}
  \bibinfo{author}{\bibnamefont{{Alexander L. Fetter}}},
  \bibinfo{journal}{Phys.l Rev. A} \textbf{\bibinfo{volume}{62}},
  \bibinfo{pages}{063617} (\bibinfo{year}{2000}).

\bibitem[{\citenamefont{Verhulst}(2005)}]{Verhulst}
\bibinfo{author}{\bibfnamefont{F.}~\bibnamefont{Verhulst}},
  \emph{\bibinfo{title}{Methods and Applications of Singular Perturbations:
  Boundary Layers and Multiple Timescale Dynamics}}
  (\bibinfo{publisher}{Springer}, \bibinfo{year}{2005}).

\bibitem[{\citenamefont{K{\"u}hnel}(2002)}]{KühnelWolfgang}
\bibinfo{author}{\bibfnamefont{W.}~\bibnamefont{K{\"u}hnel}},
  \emph{\bibinfo{title}{Differential Geometry: Curves - Surfaces - Manifolds}}
  (\bibinfo{publisher}{Amer Mathematical Society}, \bibinfo{year}{2002}).

\bibitem[{\citenamefont{Anglin}(2002)}]{PhysRevA.65.063611}
\bibinfo{author}{\bibfnamefont{J.~R.} \bibnamefont{Anglin}},
  \bibinfo{journal}{Phys. Rev. A} \textbf{\bibinfo{volume}{65}},
  \bibinfo{pages}{063611} (\bibinfo{year}{2002}).

\bibitem[{\citenamefont{Simula et~al.}(2008{\natexlab{a}})\citenamefont{Simula,
  Mizushima, and Machida}}]{PhysRevLett.101.020402}
\bibinfo{author}{\bibfnamefont{T.~P.} \bibnamefont{Simula}},
  \bibinfo{author}{\bibfnamefont{T.}~\bibnamefont{Mizushima}},
  \bibnamefont{and} \bibinfo{author}{\bibfnamefont{K.}~\bibnamefont{Machida}},
  \bibinfo{journal}{Phys. Rev. Lett.} \textbf{\bibinfo{volume}{101}},
  \bibinfo{pages}{020402} (\bibinfo{year}{2008}{\natexlab{a}}).

\bibitem[{\citenamefont{Simula et~al.}(2008{\natexlab{b}})\citenamefont{Simula,
  Mizushima, and Machida}}]{PhysRevA.78.053604}
\bibinfo{author}{\bibfnamefont{T.~P.} \bibnamefont{Simula}},
  \bibinfo{author}{\bibfnamefont{T.}~\bibnamefont{Mizushima}},
  \bibnamefont{and} \bibinfo{author}{\bibfnamefont{K.}~\bibnamefont{Machida}},
  \bibinfo{journal}{Phys. Rev. A} \textbf{\bibinfo{volume}{78}},
  \bibinfo{pages}{053604} (\bibinfo{year}{2008}{\natexlab{b}}).

\end{thebibliography}
\end{document}